\documentclass[preprint,review,12pt]{elsarticle}

\usepackage{amssymb}
\usepackage{subfigure}			
\usepackage{amsmath}
\usepackage{tabularx}
\usepackage{booktabs}				
\usepackage{natbib}
\usepackage{bm}
\usepackage{color}

\biboptions{sort&compress}

\begin{document}

\begin{frontmatter}



\title{Non-reciprocal wave propagation in discretely modulated spatiotemporal plates}


\author{E. Riva\corref{cor1}}
\author{M. Di Ronco\corref{cor2}}
\author{G. Cazzulani\corref{cor2}}
\author{F. Braghin\corref{cor2}}
\address{Politecnico di Milano\\ Department of Mechanical Engineering\\ Via La Masa 1, 20156, Milano}
\cortext[cor1]{\footnotesize{Corresponding author. Email address: \footnotesize{emanuele.riva@polimi.it}.}}

\begin{abstract}
We investigate non-reciprocal wave propagation in spatiotemporal phononic plates. 
In particular, the first goal of this manuscript is to present a general formulation of the Plane Wave Expansion Method (PWEM) that, in contrast with previous works, is applicable to any class of 2D spatiotemporal unit cells whose properties can be expanded in traveling plane waves. 
The second goal is to exploit this analysis tool in order to study a new class of materials capable of violating mirror symmetry in momentum space, therefore breaking reciprocity principle along different wave propagation directions. This is obtained by considering the plate elastic properties to be discretely modulated in space and continuously in time. 
Theoretical dispersion profiles are validated and compared with numerical simulations. 
\end{abstract}


\end{frontmatter}


\section{Introduction}
Non-reciprocal wave propagation in phononic crystals has drawn growing attention within the research community in the past years \cite{maznev2013reciprocity,zanjani2014one}. Indeed, the opportunity to design elastic structures which support one-way wave propagation can be particularly relevant for next-generation applications involving elastic energy manipulation, waveguiding and conversion \cite{nassar2017non,yi2017frequency}. \\
In this context, 1D phononic waveguides have recently been designed to break the reciprocity principle. One possibility is to locally alter material properties mimicking the propagation of a wave \cite{trainiti2016non,vila2017bloch,riva2019generalized,marconi2018observations,attarzadeh2018elastic}, which generally requires active elements. A second possibility is to leverage 2D lattice structures with spinning gyroscopes, establishing elastic analogues of the Quantum Hall Effect (QHE) \cite{wang2015topological,kane2005quantum,chen2019mechanical}. Specifically, this strategy relies on the exploitation of one-way topologically protected edge waves, thus not involving the bulk of the material. Systems with broken time reversal symmetry can be also designed using three port devices \cite{fleury2015subwavelength} or exploiting nonlinearities \cite{merkel2018dynamic,wallen2019nonreciprocal}.\\
Non-reciprocal elastic wave manipulation in 2D structures has been accomplished by Attarzadeh et al. \cite{attarzadeh2018non} leveraging spatiotemporally modulated membranes to break time reversal symmetry, continuously biasing the applied traction field and material density in space and time. 
Compared with 1D structures, the addition of one spatial dimension makes wave propagation analysis more complex, but also opens new promising ways to manipulate elastic waves along different directions. \\
Although the aforementioned examples offer a theoretical framework to study nonreciprocity in a specific class of 2D spatiotemporal mechanical structures, additional efforts are required for the analysis of arbitrarily shaped modulations. 
In pasticular, in this manuscript our aim is to bridge the gap between fully continuous modulations (for which material properties are pointwise varying in time) and those unit cell profiles that can be more easily realized in engineering applications - which must be studied with appropriate tools.\\
Motivated by this consideration, we investigate discretely modulated plates. 
This new class of materials is characterized by spatially discrete and temporally continuous elasticity profiles, mimicking the propagation of a plane wave in two dimensions. 
\color{black}
In contrast with already studied spatiotemporal materials, the proposed structure allows for a more feasible implementation, which is supposed to be provided using established techniques, such as negative capacitance shunts applied to piezoelectric materials bonded on a 2D substrate \cite{PhysRevLett.122.124301,de2008vibration}.
The corresponding non-symmetric band diagram is computed using a generalization of the Plane Wave Expansion Method (PWEM) provided in \cite{attarzadeh2018non} extending our previous work \cite{riva2019generalized} to 2D structures. The generalized PWEM can be thus applied to any kind of 2D spatiotemporal unit cells whose material properties can be projected in a Fourier basis - and not limited to any specific class of unit cell profiles - therefore assuming that the modulation can be mapped as a series of plane waves characterized by different propagation speed.\\
We demonstrate that discretely modulated plates support non-reciprocal wave propagation. Specifically, this is accomplished highlighting the leading Bloch-wave component \cite{riva2019generalized,vila2017bloch,wallen2019nonreciprocal} of the dispersion diagram, which is characterized by a wide set of supported modes. Once the diagram is filtered to the most relevant terms, associated directivity plots and group velocities are computed to predict directional wave propagation phenomena.
Theoretical dispersion plots are verified through numerical simulation of wave propagation and actual dispersion, which is reconstructed from the displacement response of a plate under tone burst excitation.\\
The article is organized as follows: in section 2 we describe the analytical procedure for the band diagram computation. In section 3, theoretical and numerical results are compared and discussed. Concluding remarks are presented in section 4.
\section{Analytical procedure for non-symmetric band diagram computation}
Consider the general equation describing the out of plane dynamics of a Kirchoff plate, therefore neglecting in-plane polarized motion:
\begin{equation}
\frac{\partial^2 m_x}{\partial x^2}+2\frac{\partial^2 m_{xy}}{\partial x \partial y}+\frac{\partial^2 m_y}{\partial y^2}=-\frac{\partial}{\partial t}\bigg[s(x,y)\rho(x,y,t)\frac{\partial w(x,y,t)}{\partial t}\bigg] 
\label{eq:01}
\end{equation}  
where $w(x,y,t)$ is the out of plane displacement field, $s(x,y)$ and $\rho(x,y,t)$ are the plate thickness and material density functions respectively. $m_{i,j}$ are the bending moment stress resultants, which read:
\begin{equation}
\begin{split}
m_x&=B(x,y,t)\bigg[\frac{\partial^2 w(x,y,t)}{\partial x^2}+\nu\frac{\partial^2 w(x,y,t)}{\partial y^2}\bigg]\\[4pt]
m_y&=B(x,y,t)\bigg[\frac{\partial^2 w(x,y,t)}{\partial y^2}+\nu\frac{\partial^2 w(x,y,t)}{\partial x^2}\bigg]\\[4pt]
m_{xy}&=(1-\nu)B(x,y,t)\frac{\partial^2 w(x,y,t)}{\partial x \partial y}
\end{split}
\label{eq:02}
\end{equation}
where $E(x,y,t)$ is the Young's Modulus and $B(x,y,t)=\frac{E(x,y,t)}{1-\nu^2}\frac{s^3(x,y)}{12}$ is the bending stiffness, which is a general function of space and time. Merging Eqs. \ref{eq:01} and \ref{eq:02} gives the PDE governing elastic wave propagation in modulated plates, reported in Appendix A for the sake of brevity. Now, under the assumption of space-time periodic elasticity and material density, $B(x,y,t)$ and $G\left(x,y,t\right)=s(x,y)\rho(x,y,t)$ can be written in terms of exponential functions:
\begin{equation}
\begin{split}
B(x,y,t)=\sum\limits_{m,n,v=-\infty}^{\infty}\hat{B}_{m,n,v}{\rm{e}}^{{\rm{j}}(\bm{\kappa_m\cdot r}-v\omega_mt)}\\
G(x,y,t)=\sum\limits_{m,n,v=-\infty}^{\infty}\hat{G}_{m,n,v}{\rm{e}}^{{\rm{j}}(\bm{\kappa_m\cdot r}-v\omega_mt)} 
\end{split}
\label{eq:03}
\end{equation}
where the term $\sum\limits_{m,n,v=-\infty}^{\infty}$ denotes a nested summation over the indexes $m,n,v$. $\bm{\kappa_m}=\left(mk_{mx},nk_{my}\right)$ and $\bm{r}=\left(x,y\right)$ are the modulation wavevector and spatial coordinates mapped within the unit cell domain. In the case at hand, we define $\kappa_{mx}=\frac{2\pi}{\lambda_{mx}}$, $\kappa_{my}=\frac{2\pi}{\lambda_{my}}$ and $\omega_m=\frac{2\pi}{T_m}$ as spatial and temporal modulation wavenumbers and frequency, with associated wavelengths $\lambda_{mx}$ and $\lambda_{my}$ and temporal period $T_m$. 
Corresponding Fourier coefficients $\hat{B}_{m,n,v}$ and $\hat{G}_{m,n,v}$ are obtained by numerical integration within the domain $D=\left[-\displaystyle\frac{\lambda_{mx}}{2},\displaystyle\frac{\lambda_{mx}}{2}\right]\times\left[-\displaystyle\frac{\lambda_{my}}{2},\frac{\lambda_{my}}{2}\right]\times\left[-\displaystyle\frac{T_m}{2},\frac{T_m}{2}\right]$:
\begin{equation}
\begin{split}
\hat{B}_{m,n,v}=\frac{1}{T_m \lambda_{mx} \lambda_{my}}\int_DB(x,y,t){\rm{e}}^{-{\rm{j}}(\bm{\kappa_m\cdot r}-v\omega_m t)}{\rm{dD}}\\[4pt]
\hat{G}_{m,n,v}=\frac{1}{T_m \lambda_{mx} \lambda_{my}}\int_DG(x,y,t){\rm{e}}^{-{\rm{j}}(\bm{\kappa_m\cdot r}-v\omega_m t)}{\rm{dD}}
\end{split}
\label{eq:04}
\end{equation}
A general formulation of the PWEM is thus presented here to compute the wave propagation properties associated with the spatiotemporal unitary cell, therefore assuming a solution $w(x,y,t)$ which is a combination of plane waves owning the same periodicity of the modulation:
\begin{equation}
w(x,y,t)=\hat{w}(x,y,t){\rm{e}}^{{\rm{j}}(\bm{k\cdot r}-\omega t)},\hspace{0.5cm}\hat{w}(x,y,t)=\sum\limits_{p,q,r=-\infty}^{\infty}\hat{W}_{p,q,r}{\rm{e}}^{{\rm{j}}(\bm{\kappa_m\cdot r}-r\omega_mt)}
\label{eq:05}
\end{equation}
where $\bm{\kappa}=\left(\kappa_x,\kappa_y\right)$ is the imposed wavevector field and $\bm{\kappa_m}=\left(pk_{mx},qk_{my}\right)$, differently from already existing - and less general - formulations in which the wave solution is reconstructed using plane waves having the same propagation speed along $x$ and $y$ directions. 
Indeed, thanks to the 3D Fourier transform of the unit cell profile, $B(x,y,t)$ and $G(x,y,t)$ are written as a series of traveling plane waves regardless the space-time function designed within the domain $D$ \cite{riva2019generalized}.
Upon combination of Eqs. \ref{eq:01}-\ref{eq:05} and enforcing orthogonality of exponential functions, the dispersion relation $\omega=\omega\left(\kappa_x,\kappa_y\right)$ yields:
\begin{equation}
\left[\bm{\tilde{L}}_0(k_x,k_y)+\bm{\tilde{L}}_1\omega+\bm{\tilde{L}}_2\omega^2\right]\bm{\tilde{w}}=0
\label{eq:06}
\end{equation}
which is a Quadratic Eigenvalue Problem (QEP) solved for $\omega$ imposing $\kappa_x$ and $\kappa_y$. $\bm{\tilde{L}}_0(k_x,k_y)$, $\bm{\tilde{L}}_1$ and $\bm{\tilde{L}}_2$ are full matrices of dimension $\Gamma_o=(2P+1)(2Q+1)(2R+1)$, being $P,Q,R$ are the Fourier expansion truncation orders. A complete description of the analytical procedure and QEP matrices is provided in Appendix A.
\section{Wave propagation in discretely modulated plates}
\begin{figure}[t!] 
	\centering
	\subfigure[]{\includegraphics[width=0.31\textwidth]{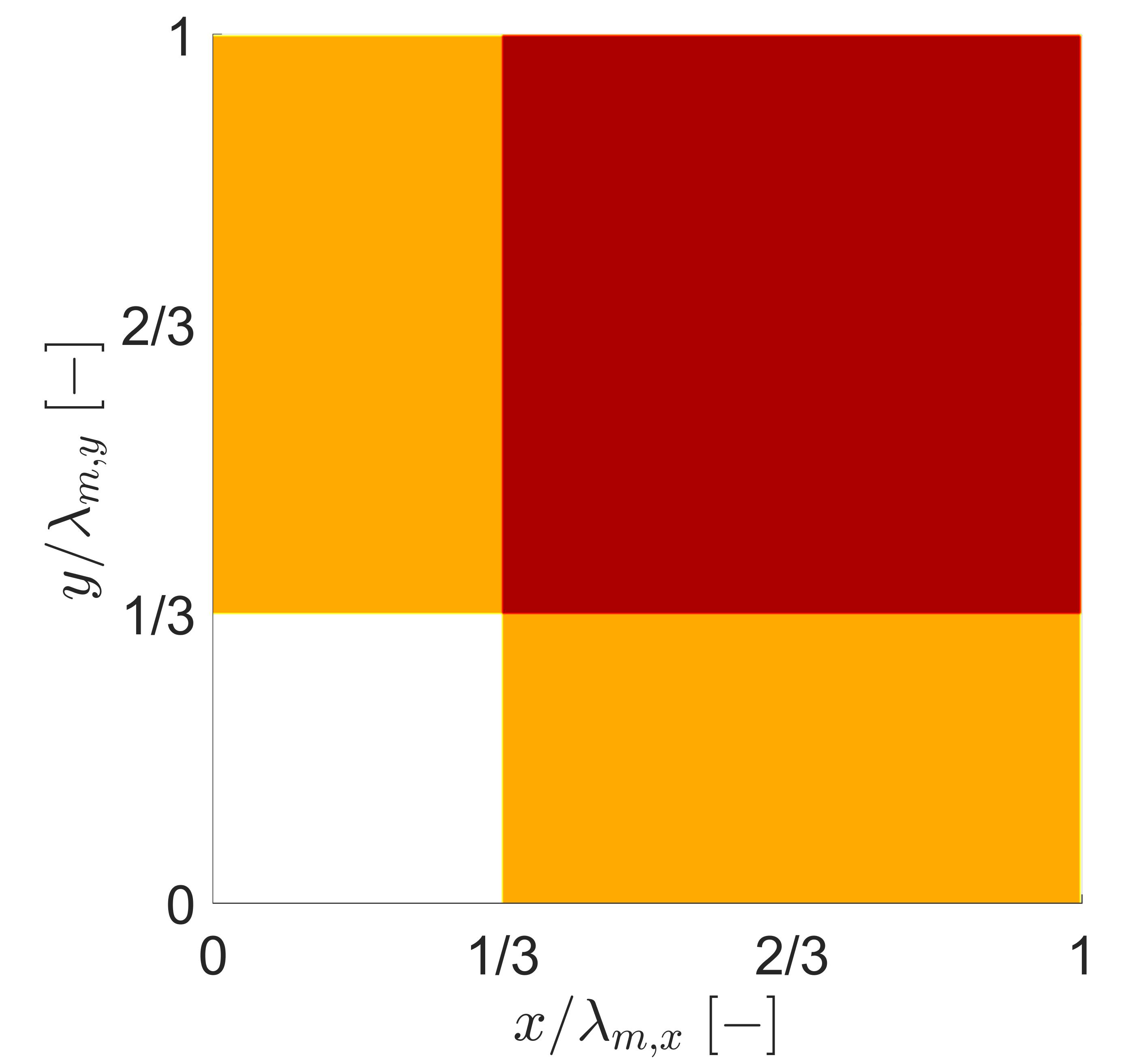} \label{fig:01_1}}\hspace{0pt}
	\subfigure[]{\includegraphics[width=0.31\textwidth]{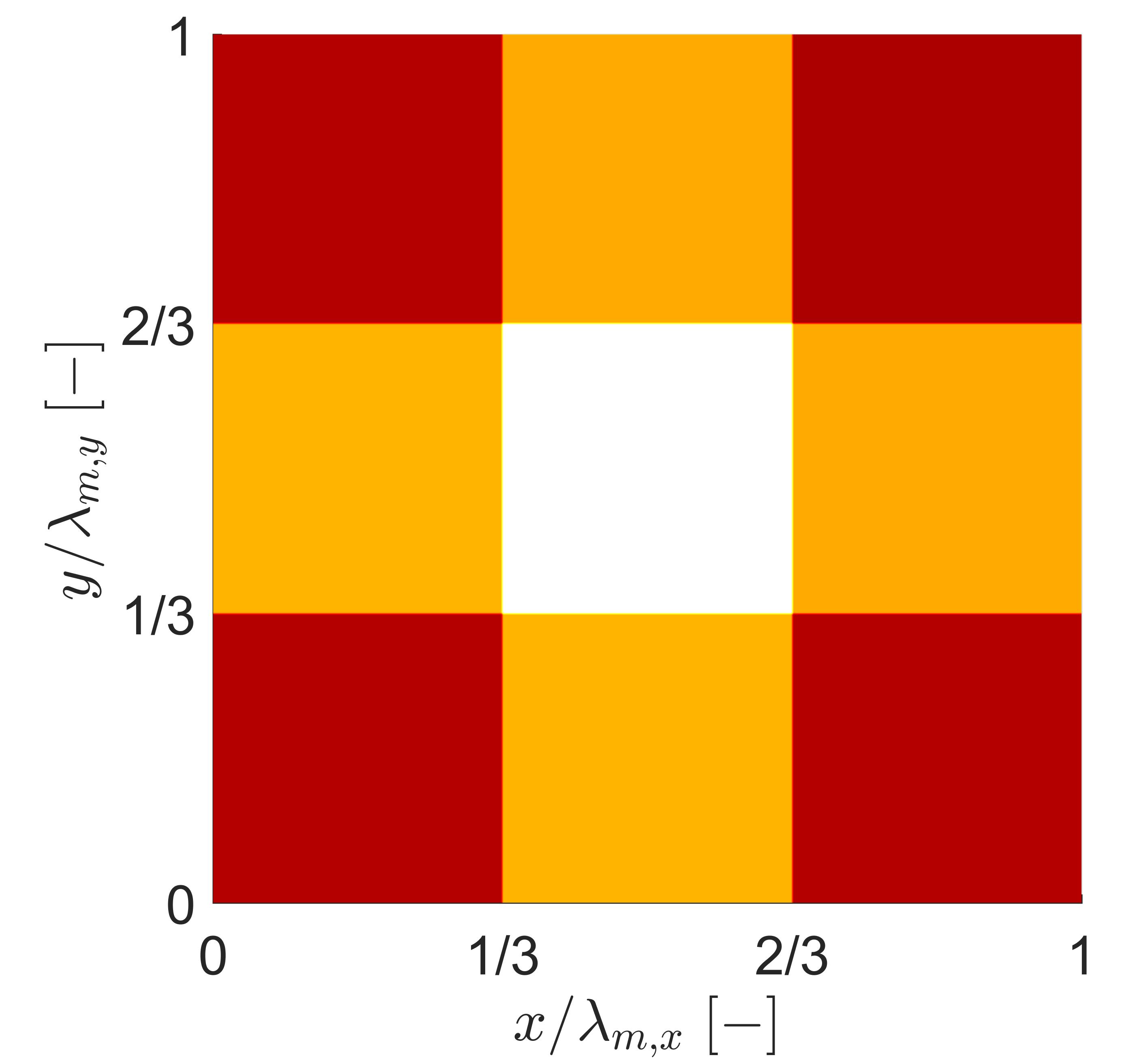} \label{fig:01_2}}\hspace{0pt}
	\subfigure[]{\includegraphics[width=0.31\textwidth]{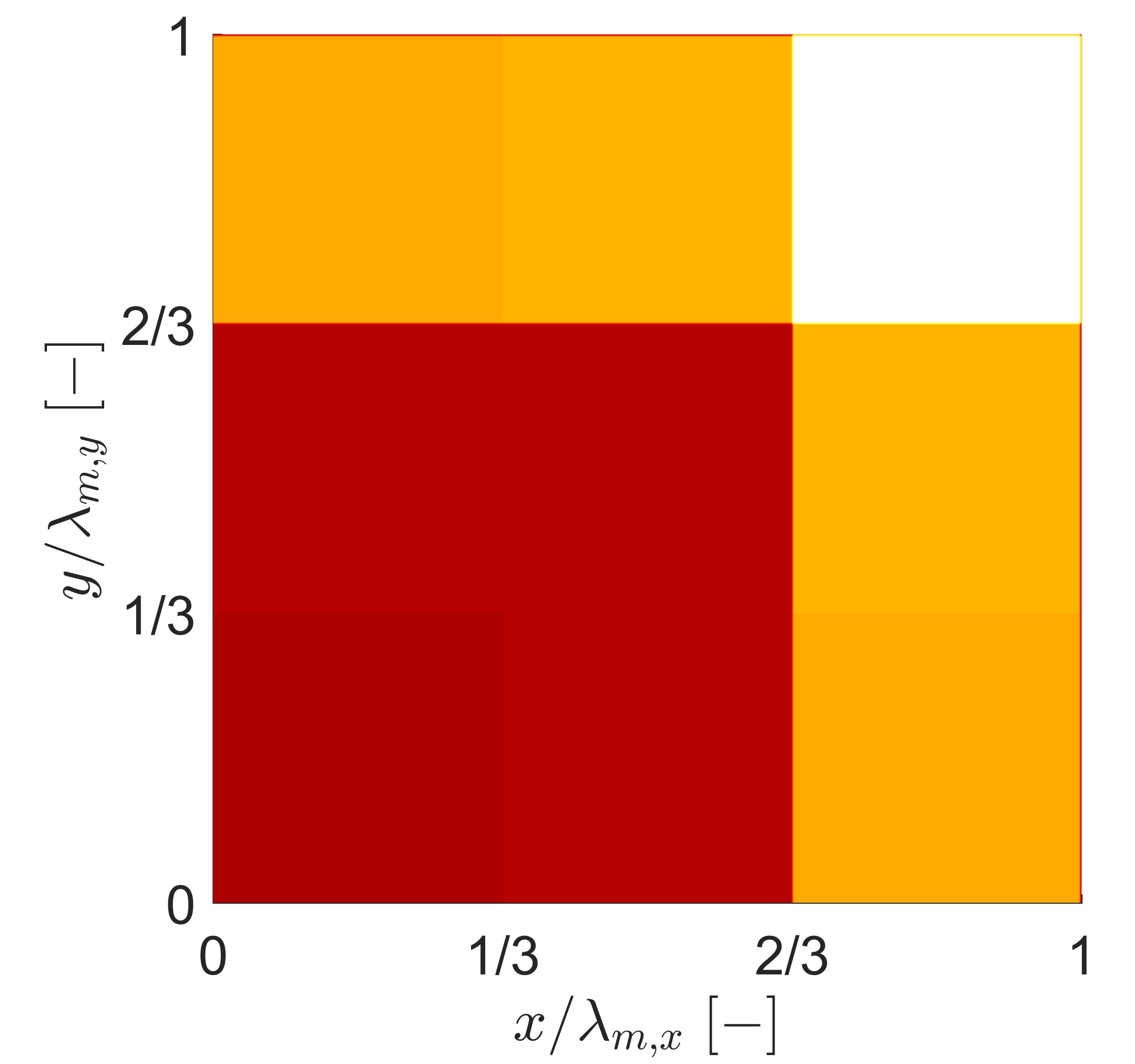} \label{fig:01_3}}
	\caption{Spatiotemporal unit cell at three different time instants. $t/T_m=0\;[-]$ \ref{fig:01_1}. $t/T_m=1/3\;[-]$ \ref{fig:01_2}. $t/T_m=2/3\;[-]$ \ref{fig:01_3}}
	\label{fig:01}
\end{figure}

\begin{figure}[!] 
	\centering
	\hspace{-33pt}\subfigure[]{\includegraphics[width=0.31\textwidth]{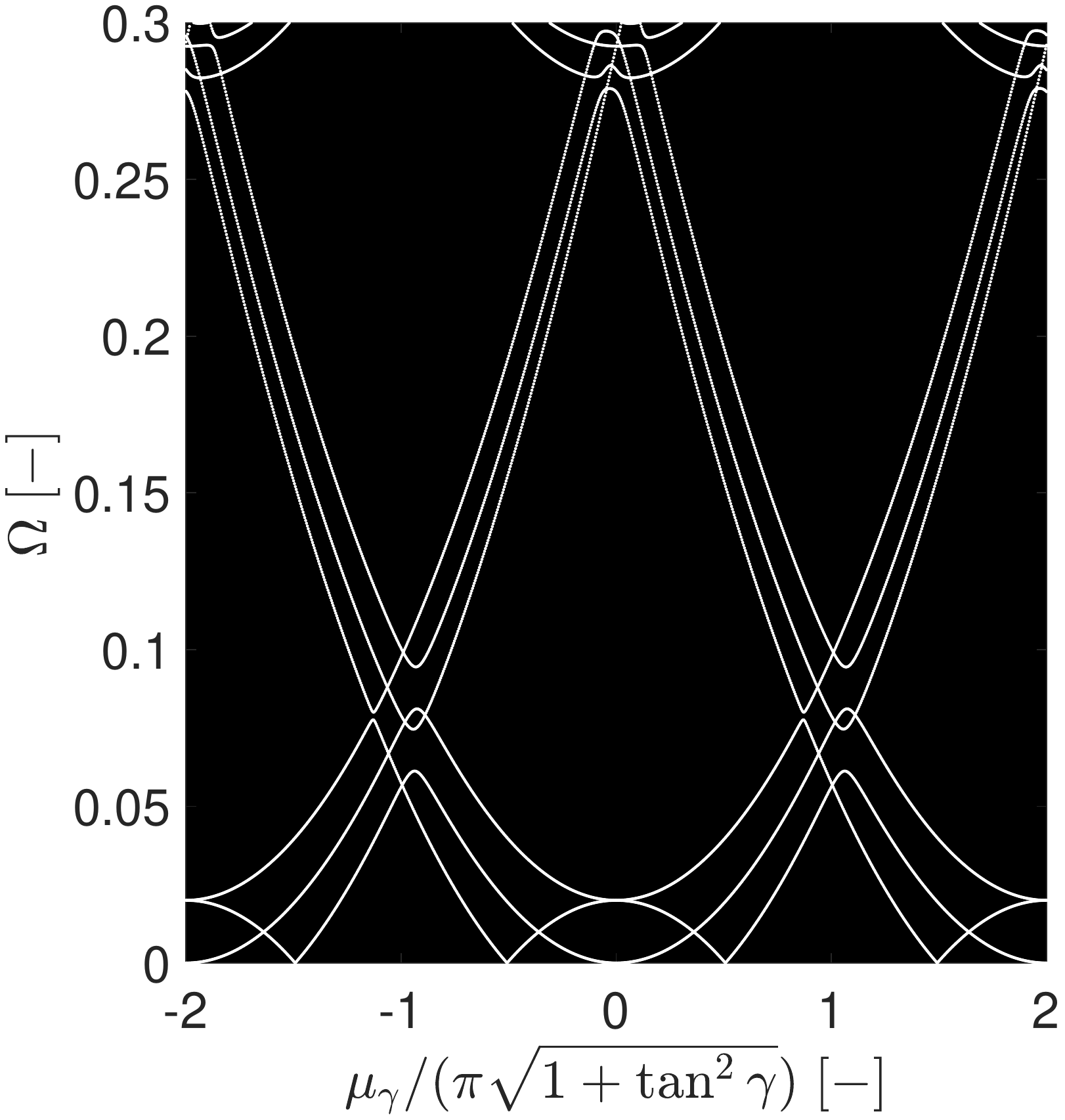} \label{fig:02_1}}\hspace{0pt}
	\subfigure[]{\includegraphics[width=0.31\textwidth]{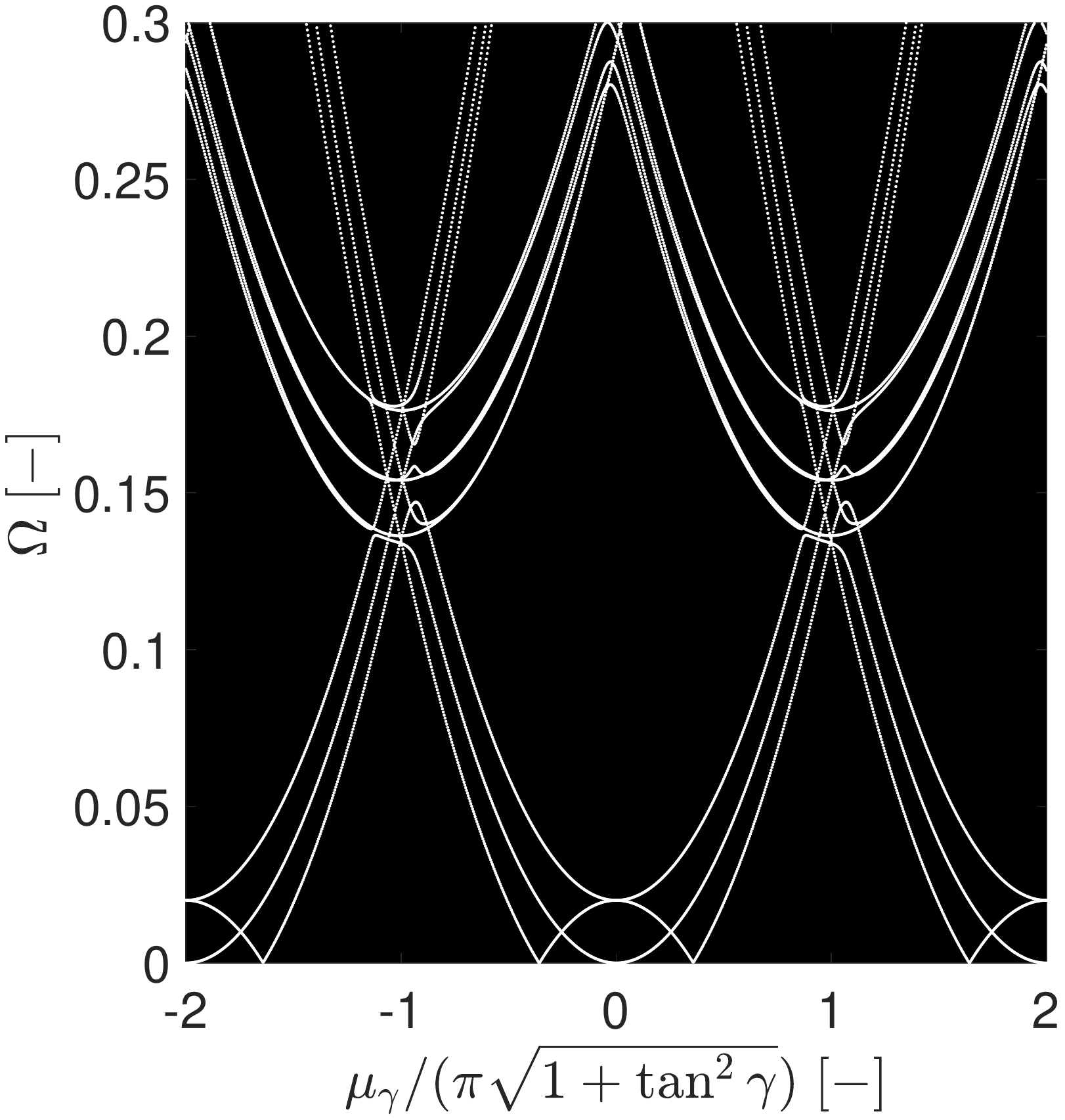} \label{fig:02_3}}
	\subfigure[]{\includegraphics[width=0.31\textwidth]{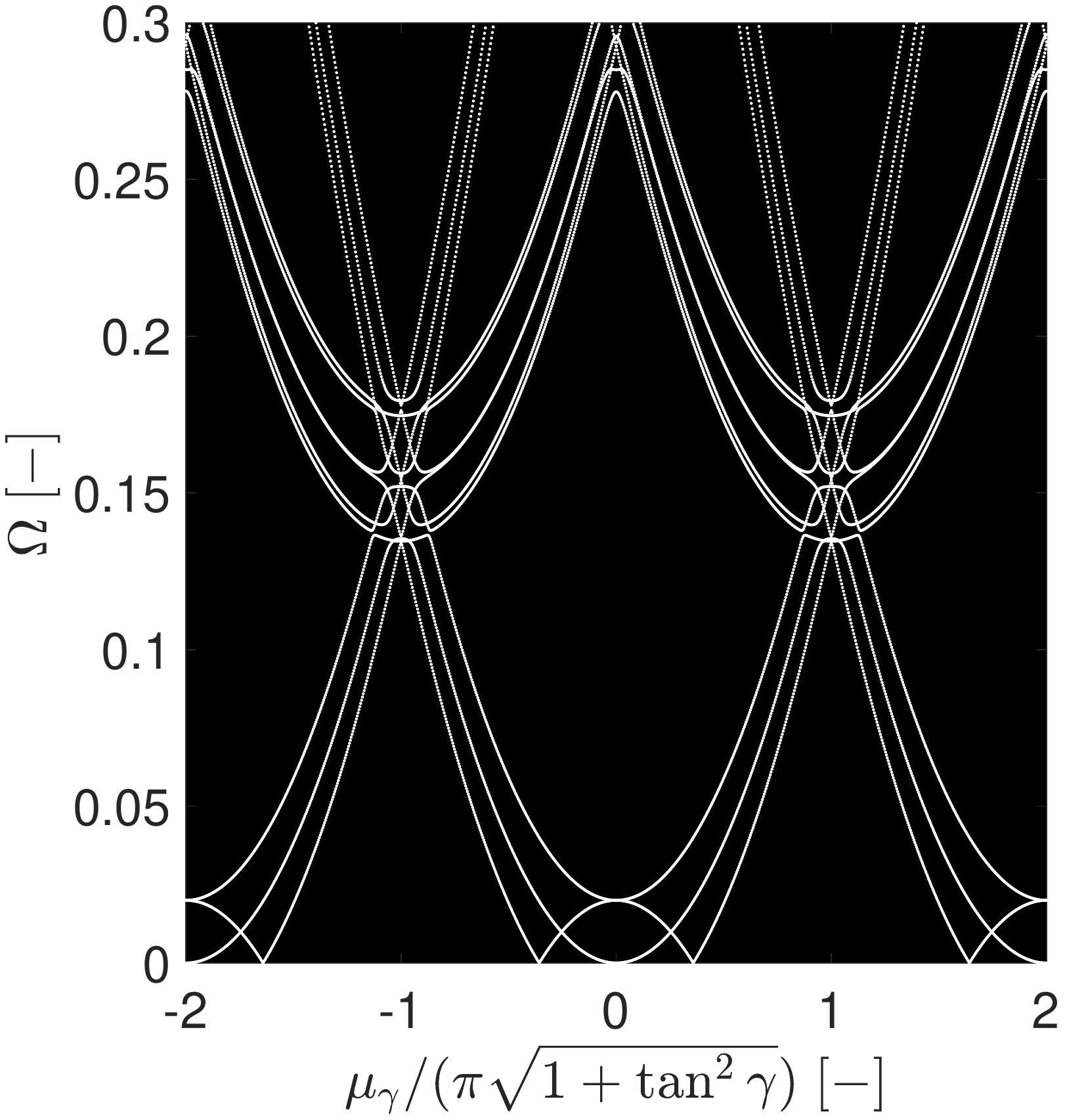} \label{fig:02_4}}\\
	\hspace{-12pt}\subfigure[]{\includegraphics[width=0.31\textwidth]{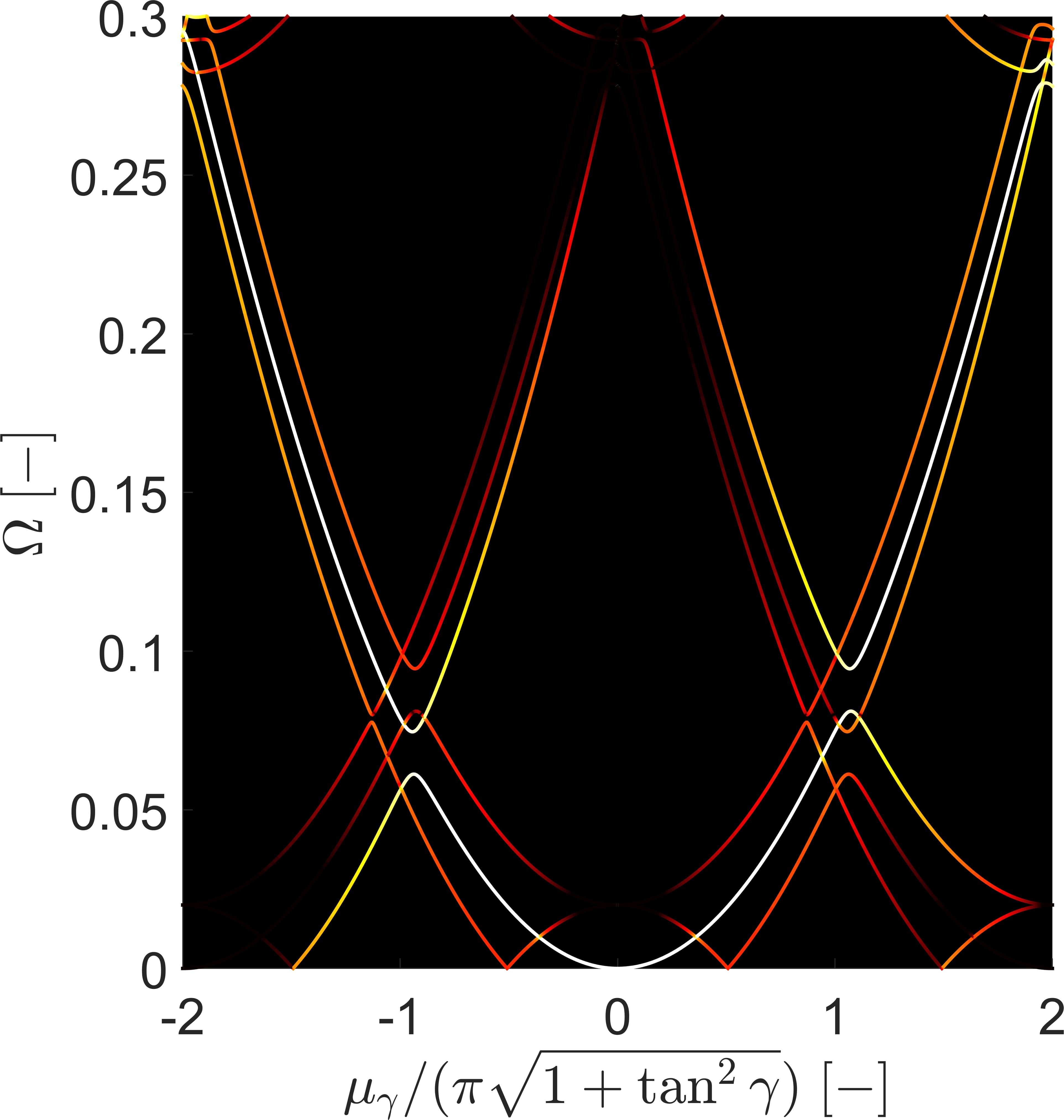} \label{fig:02_5}}
	\subfigure[]{\includegraphics[width=0.31\textwidth]{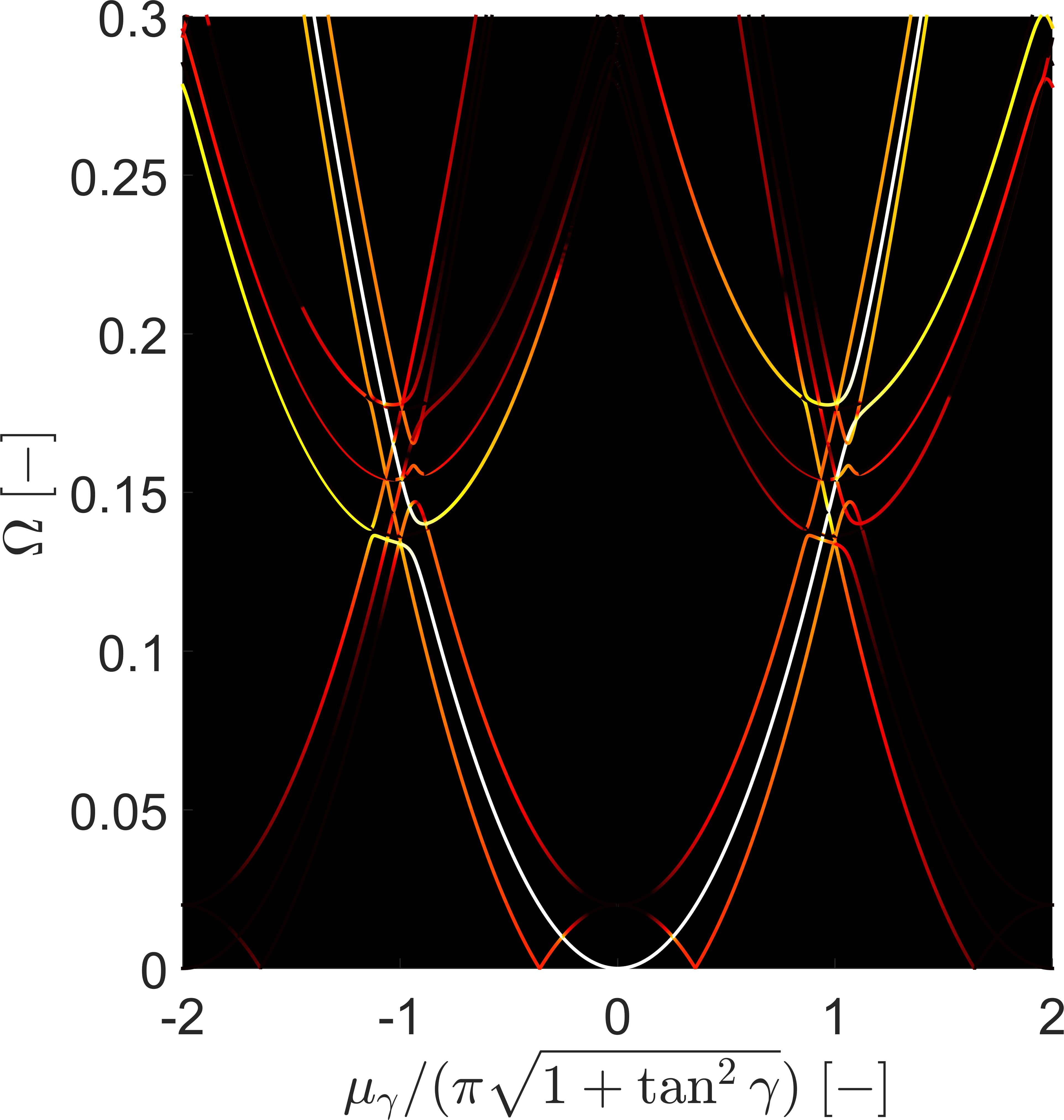} \label{fig:02_7}}
	\subfigure[]{\includegraphics[width=0.31\textwidth]{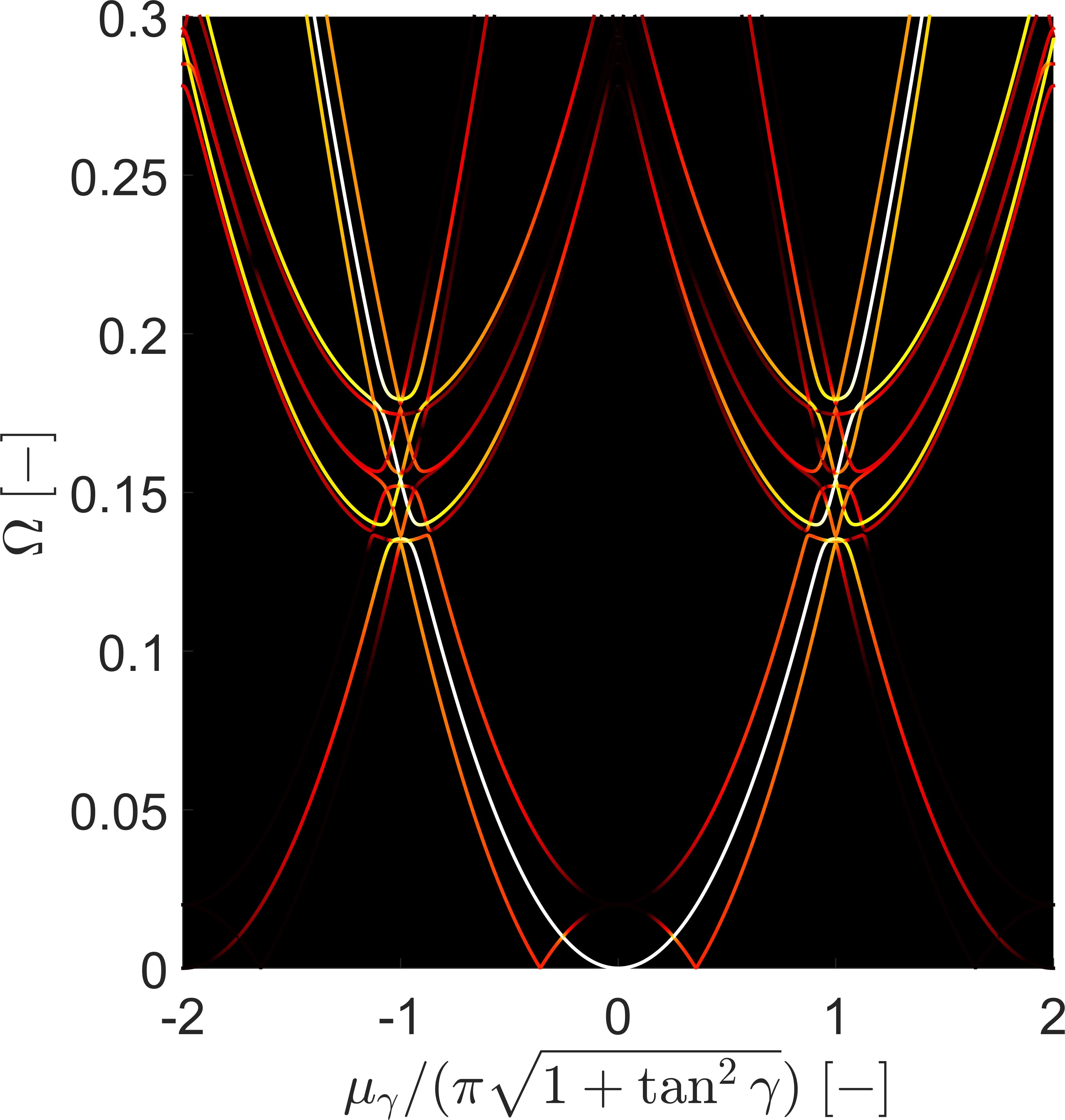} \label{fig:02_8}}	\hspace{-5pt}
	{\subfigure{\includegraphics[width=0.043\textwidth]{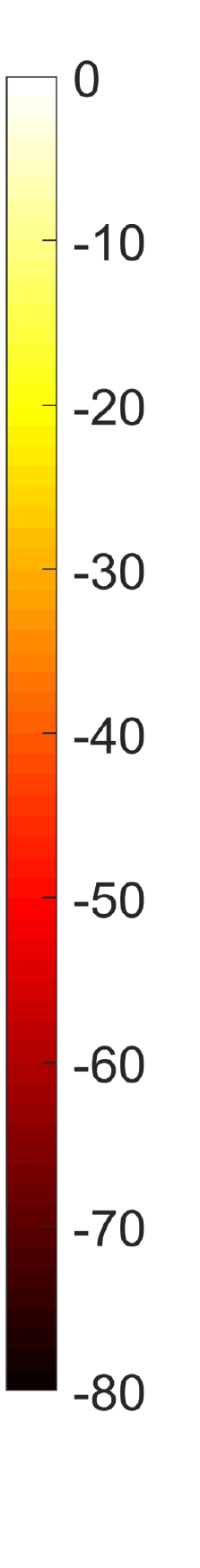}}}\\
	\hspace{-20pt} \subfigure[]{\includegraphics[width=0.31\textwidth]{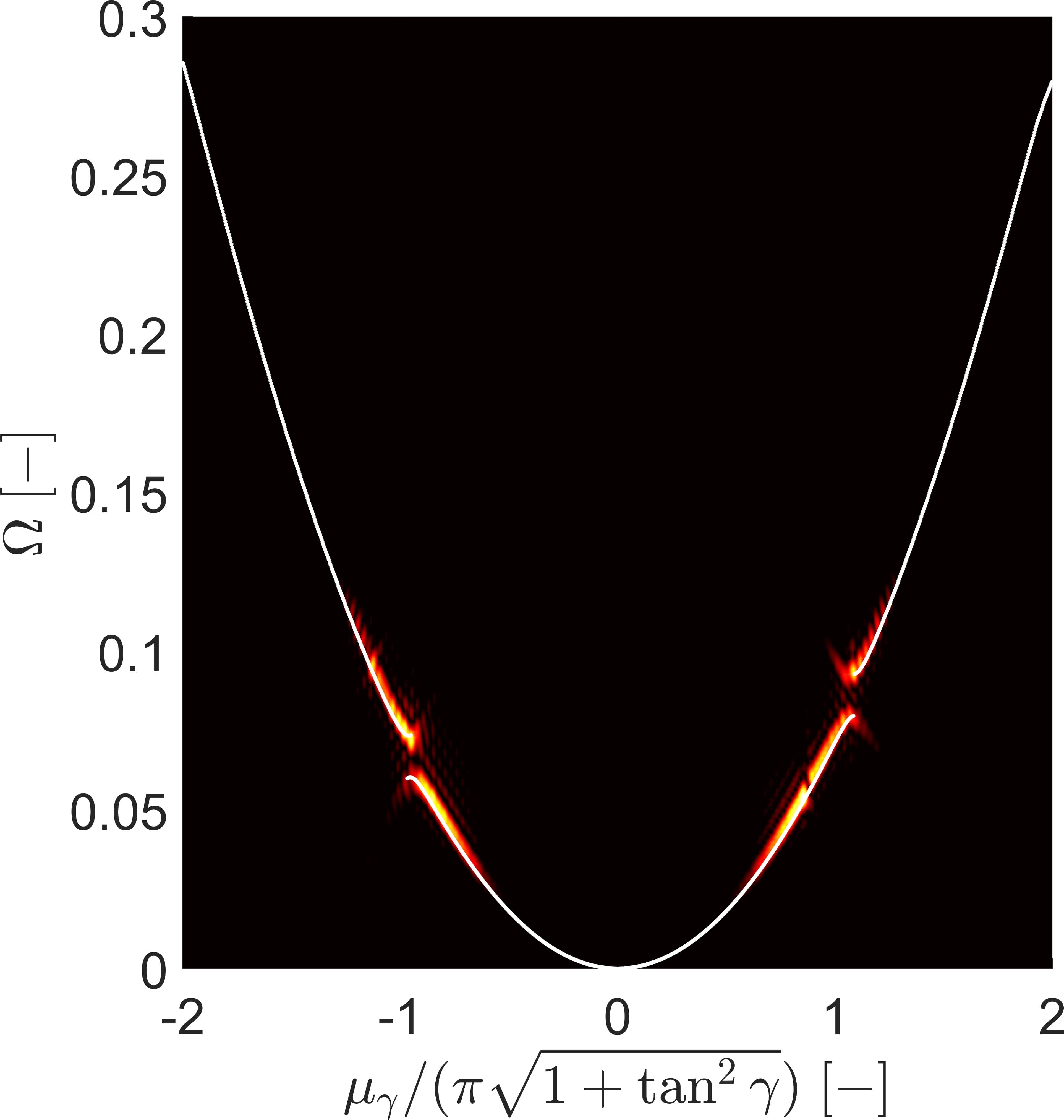} \label{fig:02_9}}\hspace{0pt}
	\subfigure[]{\includegraphics[width=0.31\textwidth]{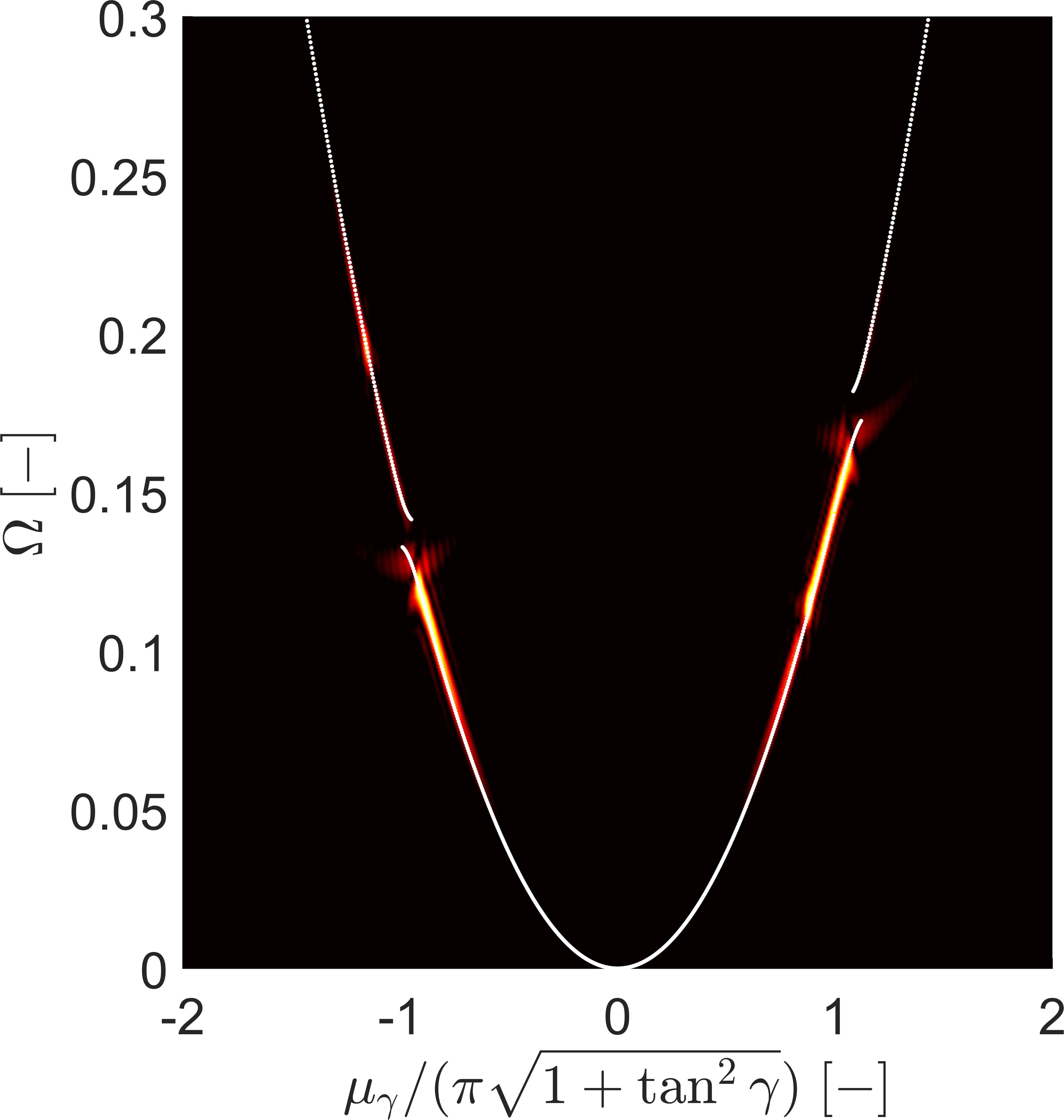} \label{fig:02_11}}
	\subfigure[]{\includegraphics[width=0.31\textwidth]{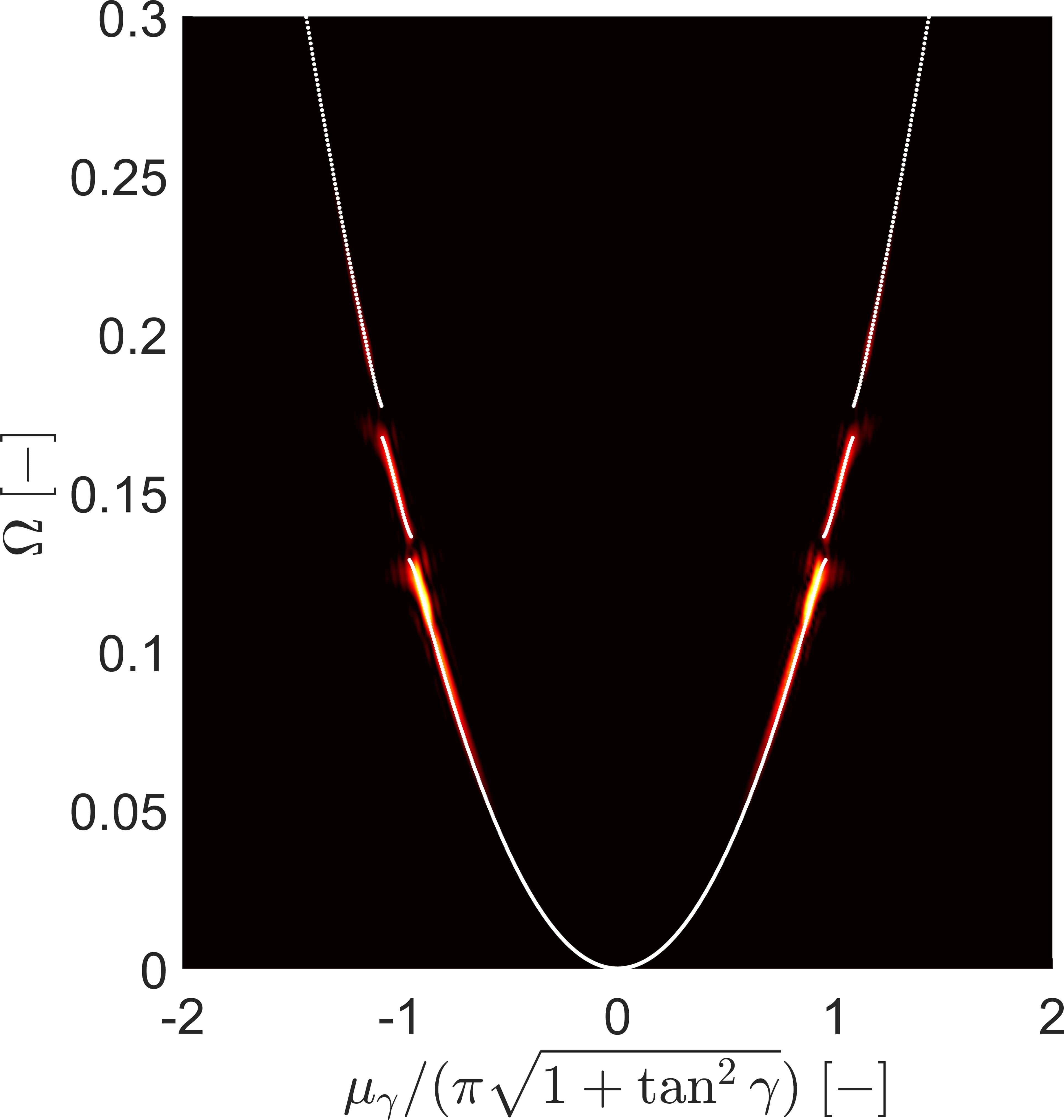} \label{fig:02_12}}	\hspace{-5pt}
	{\subfigure{\includegraphics[width=0.034\textwidth]{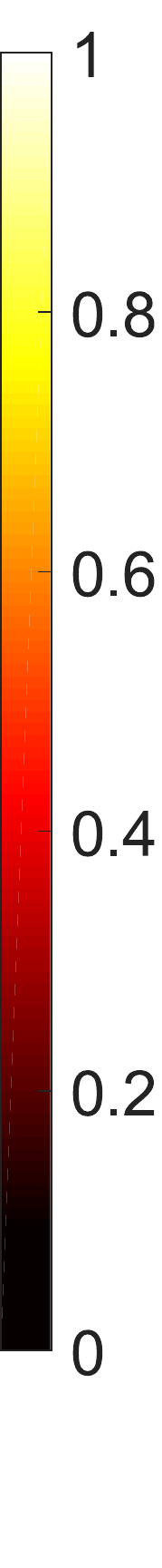}}}
	\caption{Dispersion relation $\Omega=\Omega\left(\mu_x,\mu_y\right)$ for $\gamma=0$ \ref{fig:02_1}. $\gamma=\pi/4$ \ref{fig:02_3}. $\gamma=3/4\pi$ \ref{fig:02_4}. Associated colored dispersion relation, in which the leading branches are highlighted \ref{fig:02_5}-\ref{fig:02_8} [dB]. Comparison between numerical simulations and leading dispersion branches \ref{fig:02_9}-\ref{fig:02_12}.}
	\label{fig:02}
\end{figure}

Consider a periodic plate whose elasticity is modulated according with the following law:
\begingroup\makeatletter\def\f@size{10}\check@mathfonts
\begin{equation}
\begin{split}
&E(x,y,t)=E_0\bigg\{1+\frac{\alpha_{m}}{2}\left[\cos\bigg((i-1)\frac{2\pi}{R_s}-\omega_mt\bigg)\right]+\frac{\alpha_{m}}{2}\left[\cos\bigg((j-1)\frac{2\pi}{R_s}-\omega_mt\bigg)\right]\bigg\}\quad\\[4pt]
&where:\;\;\;i=1,...,R_s\hspace{1cm}
j=1,...,R_s
\end{split}
\label{eq:07}
\end{equation}
\endgroup
where $R_s$ denotes the number of sub-cells along $x$ and $y$, $\alpha_m=E_m/E_0$ is the dimensionless modulation amplitude, thus $E_m$ and $E_0$ are the maximum and mean Young's Modulus values. $i,j$ define the temporal phase shift between consecutive sub-cells along $x$ and $y$ directions respectively. In the case at hand, the modulation is designed to mimic a plane wave that propagates along $\pi/4$ for $R_s=3$, as shown in Fig. \ref{fig:01} for three consecutive time instants. 
Plugging the Fourier expansion of $E\left(x,y,t\right)$ for $\nu=0.02$, $\alpha_{m}=0.8$ and $\lambda_{mx}=\lambda_{my}=\lambda_m=0.06\;m$ into Eq. \ref{eq:06} gives the associated dispersion relation, which is represented in Figs. \ref{fig:02_1}-\ref{fig:02_4} mapped along different wave propagation directions $\gamma$. Here $\Omega=\omega/\left(c_0\kappa_m\right)$, being $c_0=\sqrt{E_0/\rho}$ and:
\begin{equation}
\kappa_m=\frac{\kappa_{mx}\kappa_{my}}{\sqrt{\kappa_{mx}^2+\kappa_{my}^2}}\hspace{1cm}\mu_\gamma=\sqrt{\mu_x^2+\mu_y^2}=\mu_x\sqrt{1+\tan{\gamma}^2}
\end{equation}
where $\gamma=\arctan{\left(\kappa_y/\kappa_x\right)}$ and $\mu_x=\kappa_x\lambda_{mx}$.
Given the time-space periodicity of $w\left(x,y,t\right)$, the number of dispersion branches is consistent with the harmonics considered in the analysis, thus $P=3,Q=3,R=1$ and the series expansion is limited to $\left(2P+1\right)\left(2Q+1\right)$ spatial and $\left(2R+1\right)$ temporal harmonics. The leading terms are thus identified by the magnitude of the associated eigenvector components \cite{wallen2019nonreciprocal} which is shown by the colored plot in Figs. \ref{fig:02_5}-\ref{fig:02_8}, therefore revealing the presence of non-reciprocal bandgaps. In the remainder of this section we restrict the analysis to these components, thus neglecting higher order harmonics, as waves propagate mainly as their leading terms. This assumption is confirmed from numerical simulations performed using the commercial software COMSOL Multiphysics. Specifically, the system is made of $90\times90$ unit cells and forced in its central region using a wide spectrum tone burst excitation. The resulting displacement field is integrated within the spatial and temporal simulation domains to obtain the numerical Bloch diagram, proving good agreement with the analytical results, as shown in the comparison in Figs. \ref{fig:02_9}-\ref{fig:02_12} for $\gamma=0,\pi/4,3/4\pi$ respectively.
\begin{figure}[t!] 
	\centering
	\subfigure[]{\includegraphics[width=0.30\textwidth]{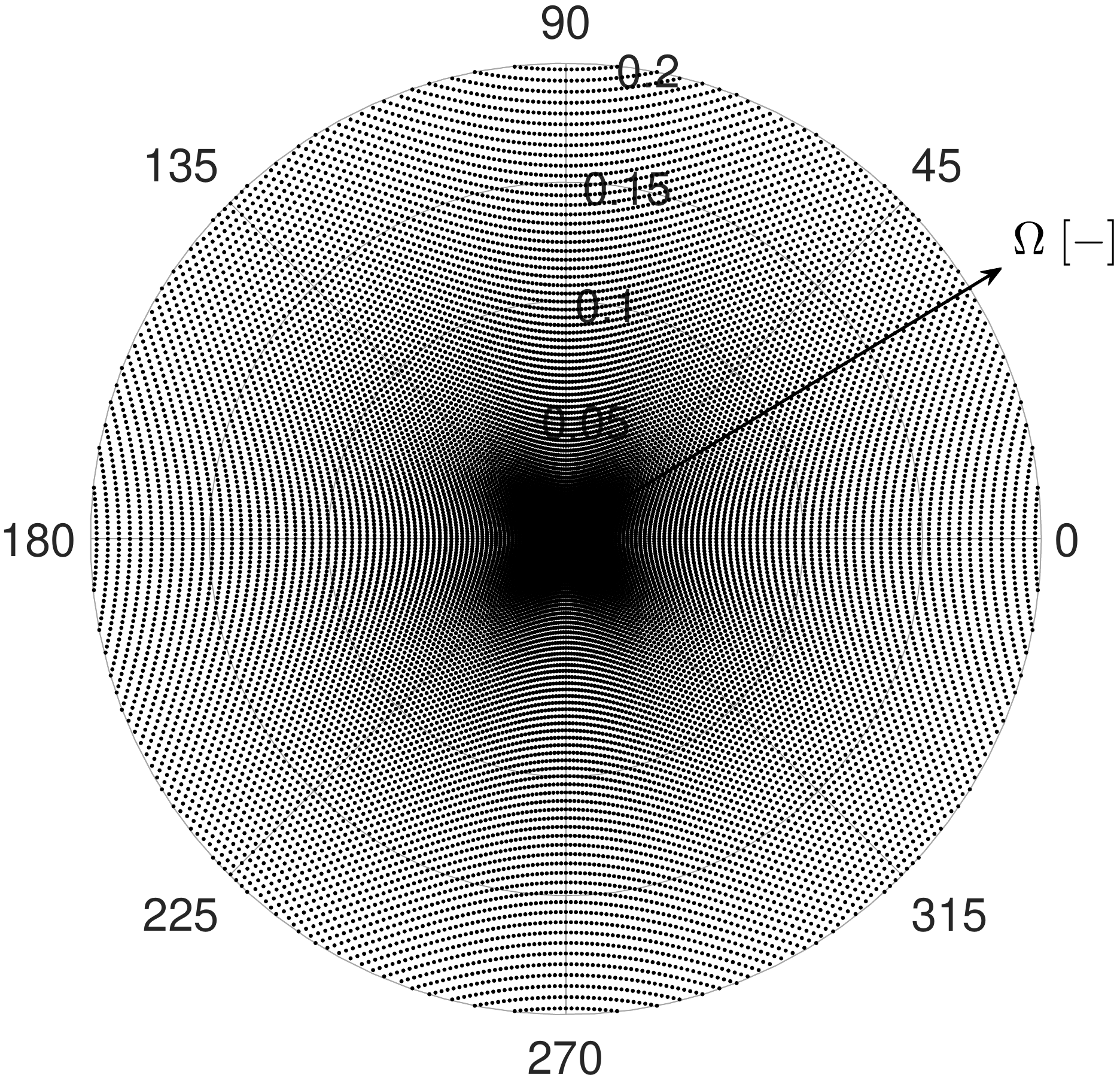} \label{fig:03_1}}\hspace{10pt}
	\subfigure[]{\includegraphics[width=0.30\textwidth]{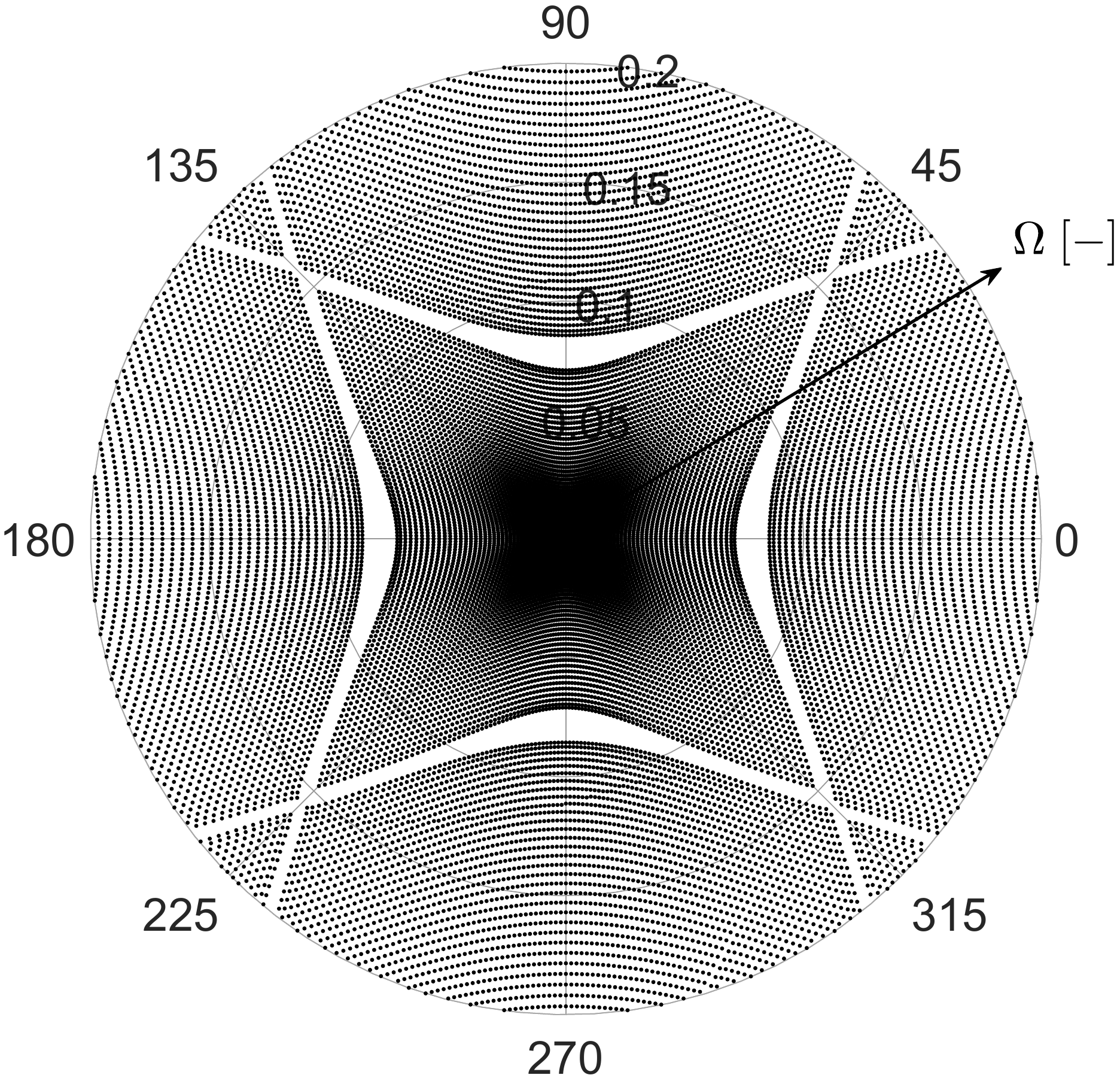} \label{fig:03_2}}\hspace{10pt}
	\subfigure[]{\includegraphics[width=0.30\textwidth]{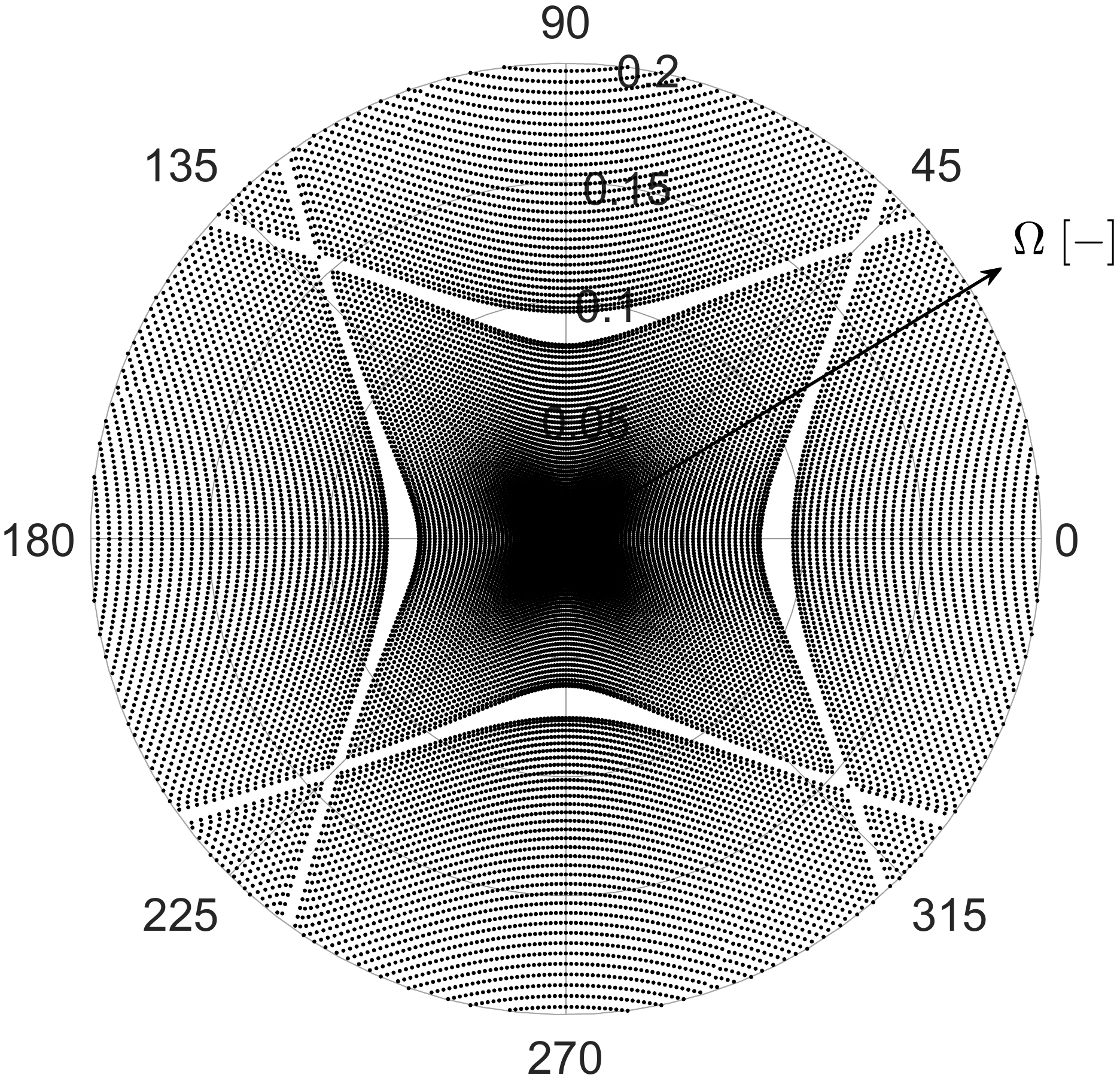} \label{fig:03_3}}\\
	\subfigure[]{\includegraphics[width=0.30\textwidth]{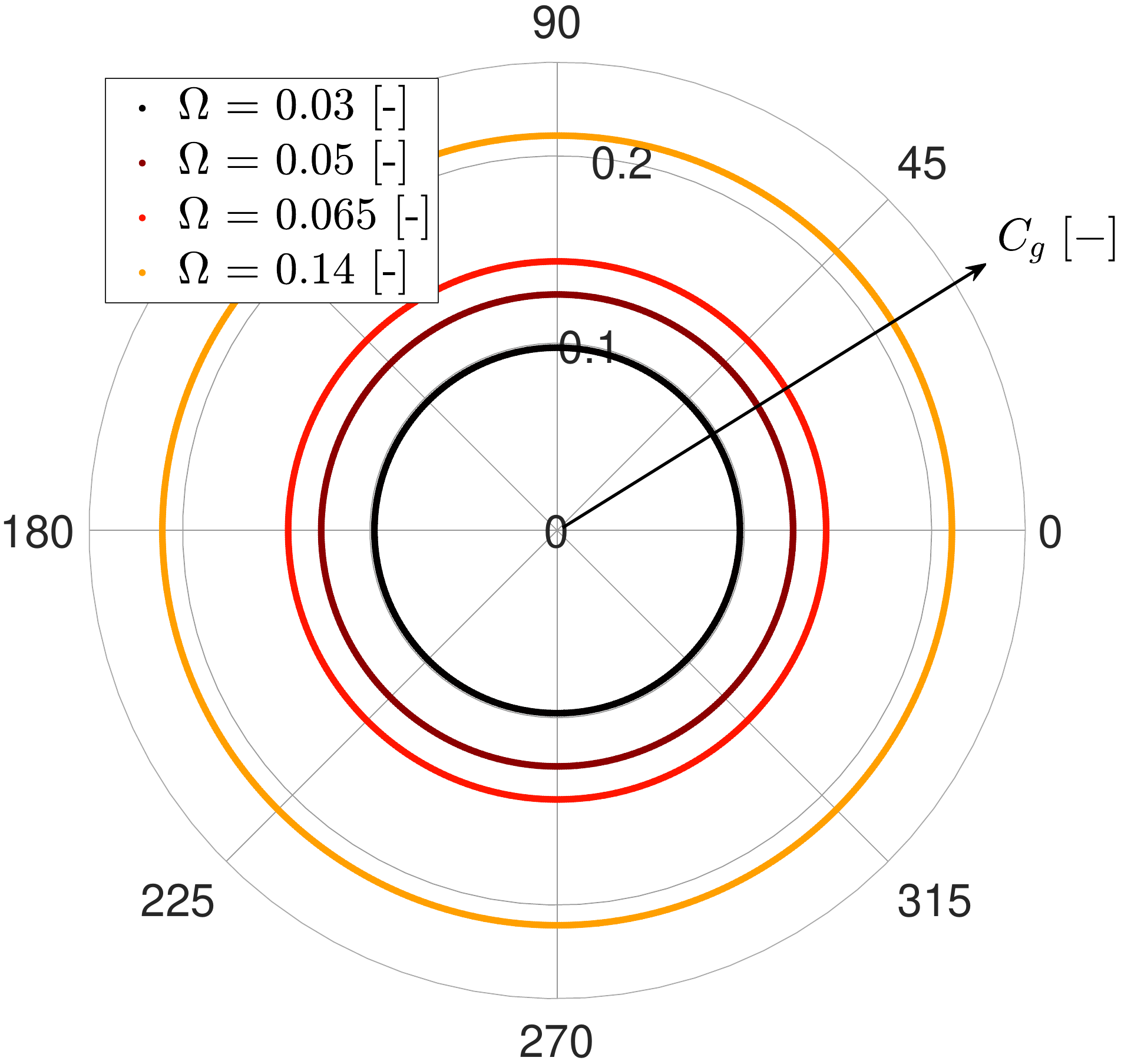} \label{fig:03_4}}\hspace{10pt}
	\subfigure[]{\includegraphics[width=0.30\textwidth]{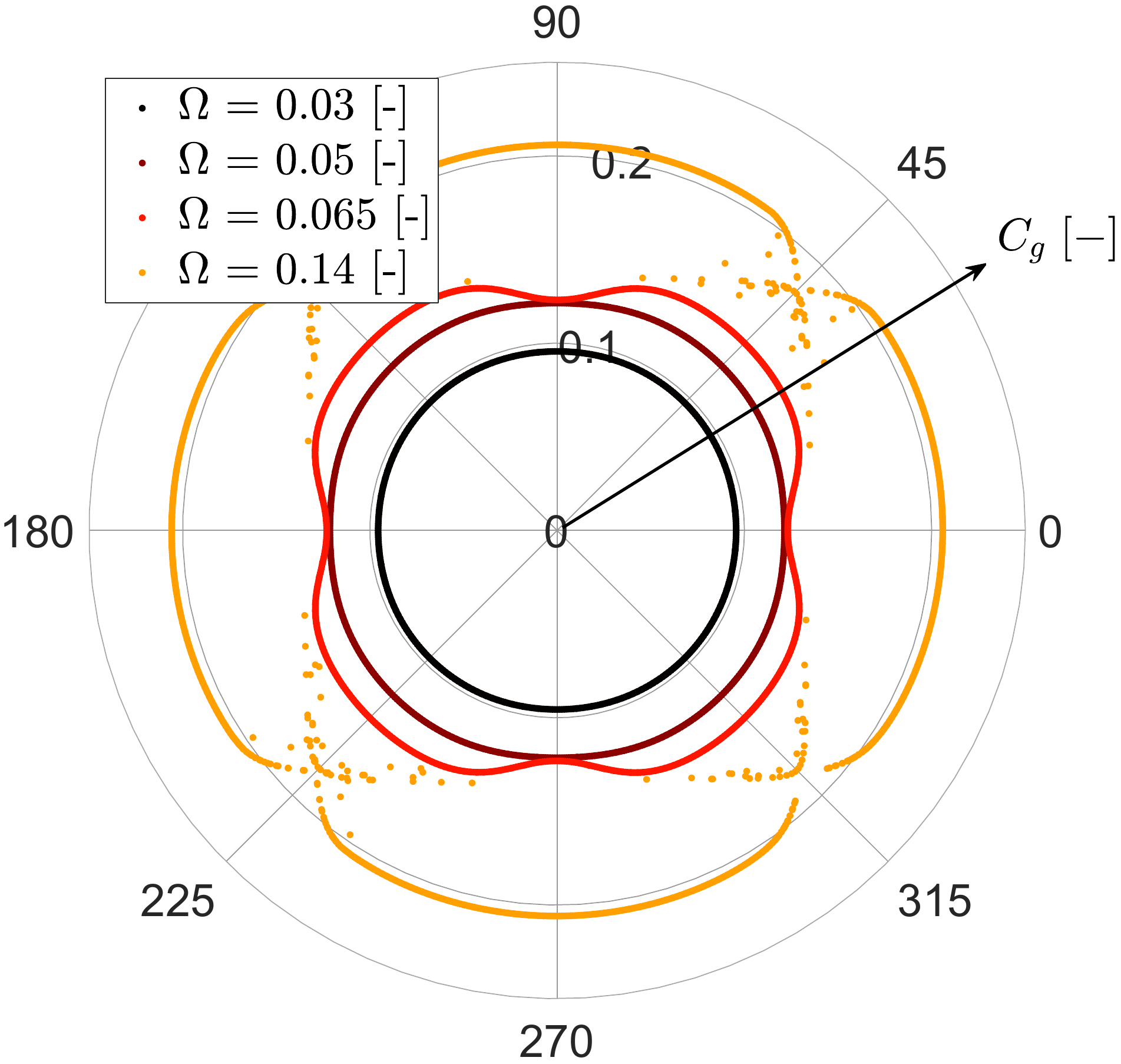} \label{fig:03_5}}\hspace{10pt}
	\subfigure[]{\includegraphics[width=0.30\textwidth]{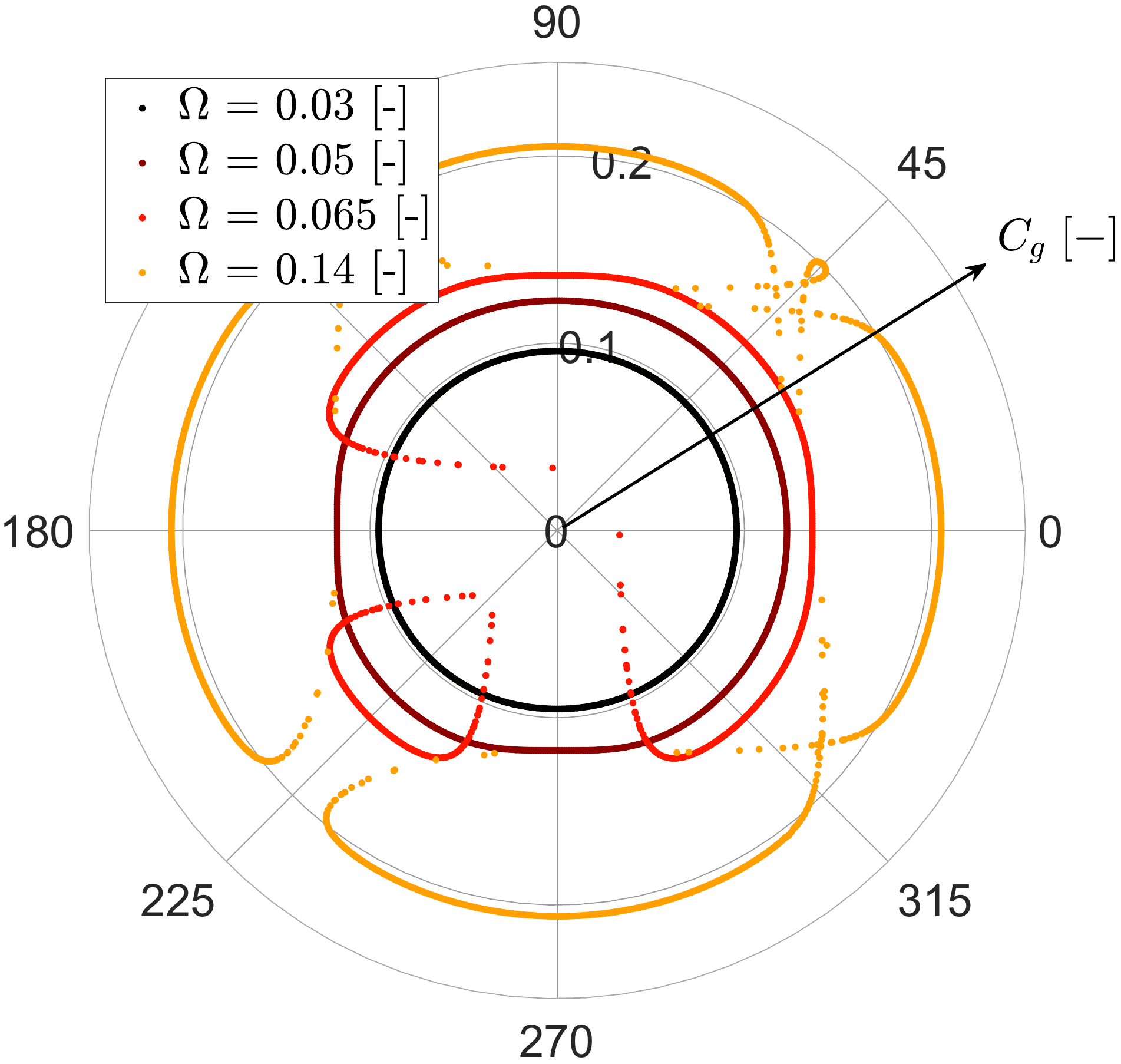} \label{fig:03_6}}
	\caption{Directivity plots for unmodulated plate $\alpha_{m}=0$, $\nu=0$ \ref{fig:03_1}, spatially modulated plate $\alpha_{m}=0.8$, $\nu=0$ \ref{fig:03_2} and spatiotemporal modulation $\alpha_{m}=0.8$, $\nu=0.02$ \ref{fig:03_3}. Associated group velocities at different frequency levels \ref{fig:03_4}-\ref{fig:03_6}. }
	\label{fig:03}
\end{figure}\\
A more complete description of the wave propagation properties is presented in Figs. \ref{fig:03_1}-\ref{fig:03_3}, in terms of directivity plot: $\Omega\left(\kappa_x,\kappa_y\right)$ is mapped in a polar plane, thus $\Omega$ is the distance from the origin and $\gamma=\arctan{\left(\kappa_y/\kappa_x\right)}$ represents the angle. Specifically, for unmodulated, spatially and spatiotemporally modulated plates. When the spatial modulation is turned on, a Bragg-bandgap opens, whose position in the frequency domain varies as a function of the direction of interest $\gamma$. The addition of a temporal periodicity results into a modulation-induced tilting of the frequency-momentum space, which is evident from the asymmetric plot in Fig. \ref{fig:03_3}. The amount of frequency bias is dictated by the scalar product between modulation velocity vector and wave propagation direction under investigation, thus:
\begin{equation}
\nu_\gamma=\bm{\nu}\cdot\frac{\bm{\kappa}}{|\bm{\kappa}|}
\end{equation}
In case of $\gamma=\pi/4$, the velocity component is $\nu_{\pi/4}=\nu_{max}$, while $\nu_{3/4\pi}=0$. As a result, band diagram is tilted along every direction but $3/4\pi+i\pi$, as already pointed out in Ref. \cite{attarzadeh2018non} for spatiotemporal membranes. 
Consider now the non-reciprocal wave propagation problem between two points A (emitter) and B (receiver). Point A is located in the central region of the plate, whereas B must be set sufficiently far from the excitation. In contrast with 1D waveguides, when the system is forced waves propagate along the direction identified by the group velocity field $C_g=\nabla\Omega\left(\mu_x,\mu_y\right)$, which is shown in Fig. \ref{fig:03} for unmodulated, spatially and spatiotemporally modulated medium. 
\begin{figure}[t!] 
	\centering
	\subfigure[]{\includegraphics[width=0.47\textwidth]{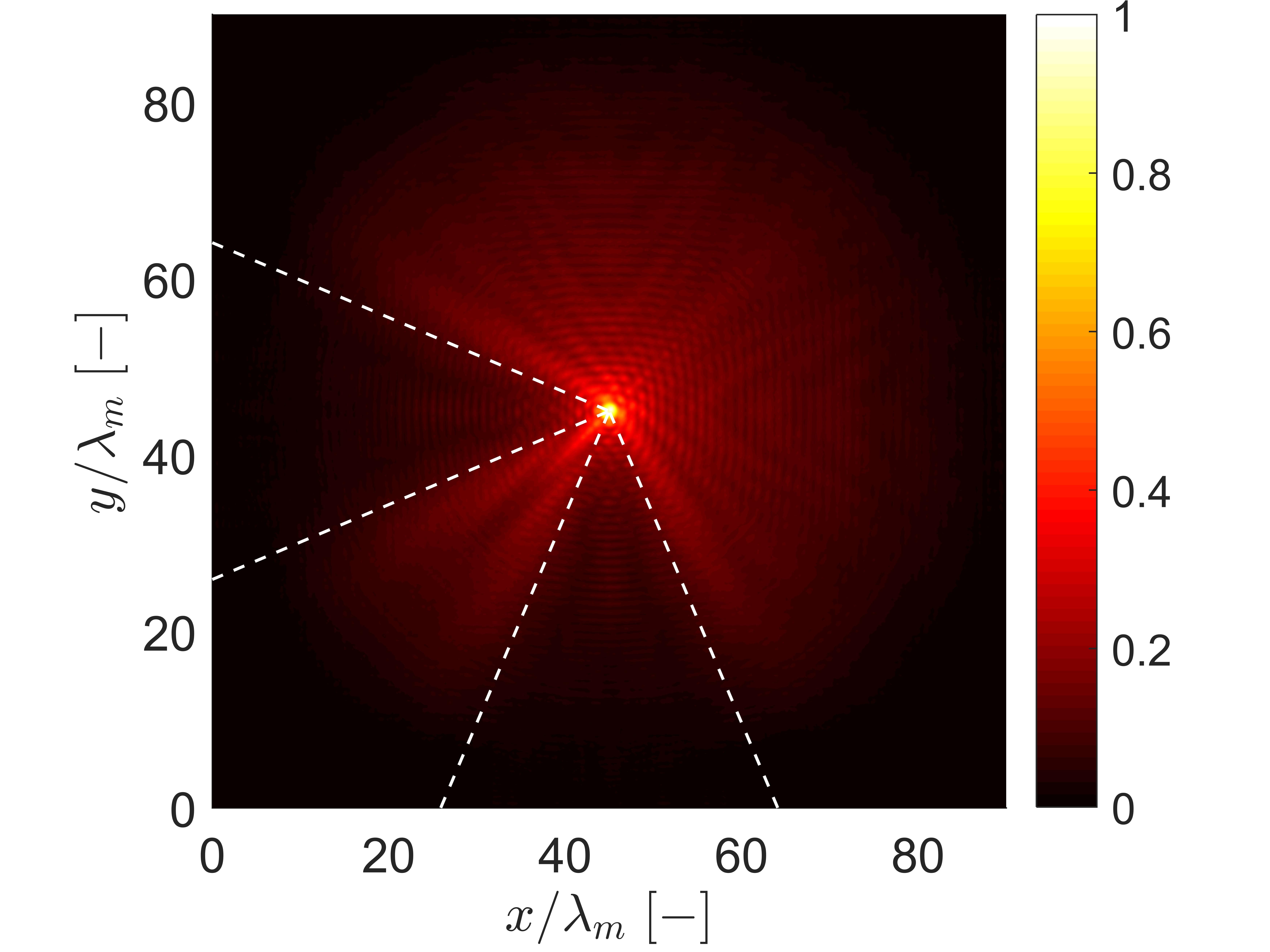} \label{fig:04_1}}\hspace{10pt}
	\subfigure[]{\includegraphics[width=0.47\textwidth]{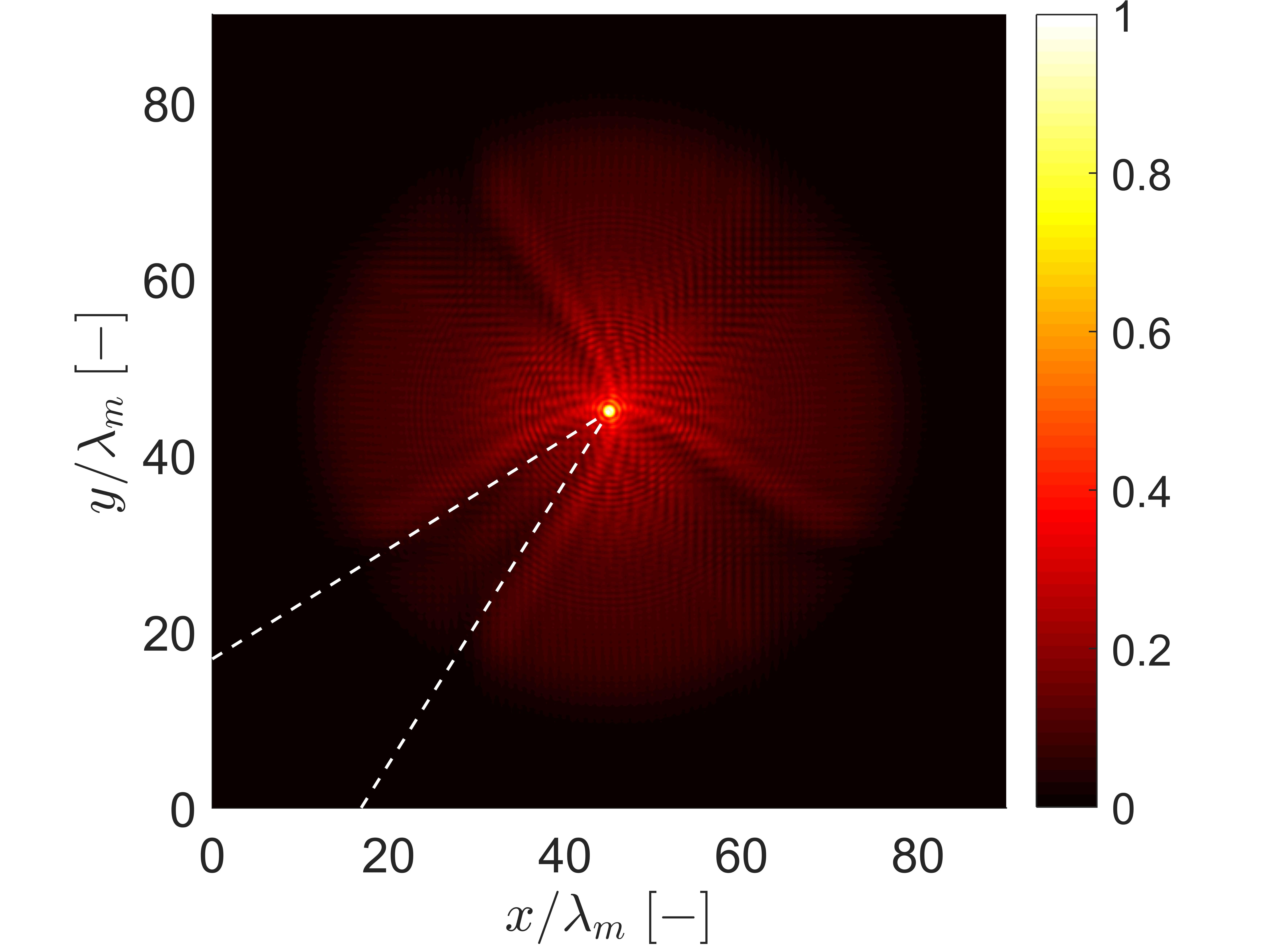} \label{fig:04_2}}
	\caption{RMS of the displacement field under narrowband tone burst excitation centered at  $\Omega=0.065$ \ref{fig:04_1} and $\Omega=0.14$ \ref{fig:04_2}.}
	\label{fig:04}
\end{figure}\\
Thanks to the spatiotemporal modulation, this latter is characterized by angular regions in which wave propagation does not occur. A non-reciprocal device can be thus achieved when the receiver B is located within $\beta_{1}\in\left[160^\circ,200^\circ\right]$ and $\beta_{2}\in\left[250^\circ,290^\circ\right]$ for $\Omega\approx0.065$, which is confirmed by the RMS of the displacement field obtained by numerical simulation under narrowband spectrum excitation as shown in Fig. \ref{fig:04_1}. The resulting input/output transmitted signal is clearly asymmetric, proving that the phononic plate is able to break time reversal symmetry along the predicted directions in Fig. \ref{fig:03_6}. Mirrored stop-regions within $\left[-20^\circ,20^\circ\right]$ and $\left[70^\circ,110^\circ\right]$ can be achieved inverting the modulation direction.\\
A similar behavior is also achieved along an angular region centered at $45^\circ$ for $\Omega\approx0.14$, i.e. along the direction of propagation of the modulation, which is tailorable, as shown by \ref{fig:03_6} and \ref{fig:04_1} where corresponding group velocity and RMS vibrational content are shown.


\section{Conclusions}
In this work we presented a general formulation of the PWEM in order to study non-reciprocal wave propagation in spatiotemporally modulated plates. This analysis tool is applicable to any class of modulations that can be written as a series of traveling plane waves, thus extending the allowable modulation classes to any type unit cell profile. Specifically, not limiting the analysis to continuous plane wave modulations, which are difficult to be realized for practical applications. \\ 
In the second part of the manuscript, the extended PWEM has been applied to spatially discrete and temporally continuous elastic medium. Numerical solutions have been compared to theoretical results, proving that the PWEM is able to correctly describe wave propagation properties of arbitrary shaped unit cell profiles. 
Specifically, we demonstrated that discretely modulated plates are able to break the reciprocity principle within different angular regions. The proposed configuration has been tested in time domain as a non-reciprocal device, which is able to support one-way elastic signals between input-output ports.

\appendix

\section{QEP matrices description}
Consider the wave propagation problem described in Section 2. The general equation governing the out of plane dynamic for a Kirchoff plate reads:
\begingroup\makeatletter\def\f@size{8}\check@mathfonts\begin{gather}
B\bigg[\frac{\partial^4 w}{\partial x^4}+2\frac{\partial^4 w}{\partial x^2 \partial y^2}+\frac{\partial^4 w}{\partial y^4}\bigg]+2\frac{\partial B}{\partial x}\frac{\partial}{\partial x}\bigg[\frac{\partial^2 w}{\partial x^2}+\frac{\partial^2 w}{\partial y^2}\bigg]+2\frac{\partial B}{\partial y}\frac{\partial}{\partial y}\bigg[\frac{\partial^2 w}{\partial x^2}+\frac{\partial^2 w}{\partial y^2}\bigg] +\notag\\
+\bigg[\frac{\partial^2 B}{\partial x^2}+\frac{\partial^2 B}{\partial y^2}\bigg]\bigg[\frac{\partial^2 w}{\partial x^2}+\frac{\partial^2 w}{\partial y^2}\bigg]-(1-\nu)\bigg[\frac{\partial^2 B}{\partial x^2}\frac{\partial^2 w}{\partial y^2}-2\frac{\partial^2 B}{\partial x \partial y}\frac{\partial^2 w}{\partial x \partial y}+\frac{\partial^2 B}{\partial y^2}\frac{\partial^2 w}{\partial x^2}\bigg]=\notag\\
=-\frac{\partial G}{\partial t}\frac{\partial w}{\partial t}-G\frac{\partial^2 w}{\partial t^2}
\label{eq6}
\end{gather}
\endgroup
Plugging Eqs. \ref{eq:03}-\ref{eq:05} into Eq. \ref{eq6} gives:
\begingroup\makeatletter\def\f@size{8}\check@mathfonts
\begin{gather}
\sum\limits_{m,n,v,p,q,r}\hat{B}_{m,n,v}\hat{W}_{p,q,r}\bigg\{\bigg[(pk_{mx}+k_x)^2+(qk_{my}+k_y)^2\bigg]\bigg[([m+p]k_{mx}+k_x)^2+([n+q]k_{my}+k_y)^2\bigg]\notag+\\
-(1-\nu)\bigg[(mk_{mx})(qk_{my}+k_y)-(nk_{my})(pk_{mx}+k_x)\bigg]^2\bigg\}\cdot {\rm{e}}^{{\rm{j}}([m+p]k_{mx}x+[n+q]k_{my}y-[v+r]\omega_mt)}=\notag\\[4pt]
=\sum\limits_{m,n,v,p,q,r}\hat{G}_{m,n,v}\hat{W}_{p,q,r}\bigg\{(r\omega_m+\omega)([v+r]\omega_m+\omega)\bigg\}\cdot {\rm{e}}^{{\rm{j}}([m+p]k_{mx}x+[n+q]k_{my}y-[v+r]\omega_mt)}\notag\\
\label{eq18}
\end{gather}
\endgroup
which can be simplified exploiting the orthogonality of the Fourier basis, thus all the terms are multiplied by ${\rm{e}}^{{-\rm{j}}(ak_{mx}x+bk_{my}y-c\omega_mt)}$ and integrated over $D=\left[-\frac{\lambda_{mx}}{2},\frac{\lambda_{mx}}{2}\right]\times\left[-\frac{\lambda_{my}}{2},\frac{\lambda_{my}}{2}\right]\times\left[-\frac{T_m}{2},\frac{T_m}{2}\right]$. Eq. \eqref{eq18}, can now be conveniently rewritten as:
\begingroup\makeatletter\def\f@size{8}\check@mathfonts
\begin{gather}
\sum\limits_{p=-P}^{P}\sum\limits_{q=-Q}^{Q}\sum\limits_{r=-R}^{R}\hat{B}_{a-p,b-q,c-r}\hat{W}_{p,q,r}\bigg\{\bigg[(pk_{mx}+k_x)^2+(qk_{my}+k_y)^2\bigg]\bigg[(ak_{mx}+k_x)^2+(bk_{my}+k_y)^2\bigg]\notag+\\
-(1-\nu)\bigg[([a-p]k_{mx})(qk_{my}+k_y)-([b-q]k_{my})(pk_{mx}+k_x)\bigg]^2\bigg\}=\notag\\
=\sum\limits_{p=-P}^{P}\sum\limits_{q=-Q}^{Q}\sum\limits_{r=-R}^{R}\hat{G}_{a-p,b-q,c-r}\hat{W}_{p,q,r}\bigg\{(r\omega_m+\omega)(c\omega_m+\omega)\bigg\}
\label{eq19}
\end{gather}
\endgroup
Eq. \ref{eq19} is a QEP, which can be written using a compact matrix notation, expanding the inner summation terms:
\begingroup\makeatletter\def\f@size{8}\check@mathfonts
\begin{gather}
\sum\limits_{p=-P}^{P}\sum\limits_{q=-Q}^{Q}\bm{\tilde{K}}_{a-p,b-q}\bigg\{\bigg[(pk_{mx}+k_x)^2+(qk_{my}+k_y)^2\bigg]\bigg[(ak_{mx}+k_x)^2+(bk_{my}+k_y)^2\bigg]\notag+\\
-(1-\nu)\bigg[([a-p]k_{mx})(qk_{my}+k_y)-([b-q]k_{my})(pk_{mx}+k_x)\bigg]^2\bigg\}\bm{\tilde{w}}_{p,q}=\notag\\
=\sum\limits_{p=-P}^{P}\sum\limits_{q=-Q}^{Q}\bigg\{\bm{\tilde{M}}^0_{a-p,b-q}+\bm{\tilde{M}}^1_{a-p,b-q}\omega+\bm{\tilde{M}}^2_{a-p,b-q}\omega^2\bigg\}\bm{\tilde{w}}_{p,q}\label{A2}
\end{gather}
\endgroup
where $\bm{\tilde{K}}_{a-p,b-q}$, $\bm{\tilde{M}}^0_{a-p,b-q}$, $\bm{\tilde{M}}^1_{a-p,b-q}$, $\bm{\tilde{M}}^2_{a-p,b-q}$ are full square matrices of size $(2R+1)$:
\begingroup\makeatletter\def\f@size{8}\check@mathfonts
\begin{align}
\bm{\tilde{K}}_{a-p,b-q}=&
\begin{bmatrix}
\hat{B}_{a-p,b-q,0} & \dots & \hat{B}_{a-p,b-q,-2R} \\
\vdots & \ddots & \vdots \\
\hat{B}_{a-p,b-q,2R} & \dots & \hat{B}_{a-p,b-q,0}
\end{bmatrix} \notag\\[0.2cm]
\bm{\tilde{M}}^0_{a-p,b-q}=&
\begin{bmatrix}
\hat{G}_{a-p,b-q,0}(-R)(-R) & \dots & \hat{G}_{a-p,b-q,-2R}(-R)(+R) \\
\vdots & \ddots & \vdots \\
\hat{G}_{a-p,b-q,2R}(+R)(-R) & \dots & \hat{G}_{a-p,b-q,0}(+R)(+R)
\end{bmatrix}\omega_m^2 \notag\\[0.2cm]
\bm{\tilde{M}}^1_{a-p,b-q}=&
\begin{bmatrix}
\hat{G}_{a-p,b-q,0}(-R-R) & \dots & \hat{G}_{a-p,b-q,-2R}(-R+R) \\
\vdots & \ddots & \vdots \\
\hat{G}_{a-p,b-q,2R}(+R-R) & \dots & \hat{G}_{a-p,b-q,0}(+R+R)
\end{bmatrix}\omega_m \notag\\[0.2cm]
\bm{\tilde{M}}^2_{a-p,b-q}=&
\begin{bmatrix}
\hat{G}_{a-p,b-q,0} & \dots & \hat{G}_{a-p,b-q,-2R} \\
\vdots & \ddots & \vdots \\
\hat{G}_{a-p,b-q,2R} & \dots & \hat{G}_{a-p,b-q,0}
\end{bmatrix} \notag
\end{align}
\endgroup
and $\bm{\tilde{w}}_{p,q}$ accommodates the $(2R+1)$ time-harmonic components:
\begin{equation}
\bm{\tilde{w}}_{p,q}=\{w_{p,q,-R},\dots ,w_{p,q,+R}\}^T\notag
\end{equation}
In the same way, the summation term $q\in[-Q,Q]$ can be expanded, thus:
\begingroup\makeatletter\def\f@size{8}\check@mathfonts
\begin{equation}
\sum\limits_{p=-P}^{P}\bigg(\bm{\tilde{Z}}^0_{a-p}+\bm{\tilde{Z}}^1_{a-p}\omega+\bm{\tilde{Z}}^2_{a-p}\omega^2\bigg)\bm{\tilde{w}}_{p}=0
\end{equation}
\endgroup
where $\bm{\tilde{Z}}^0_{a-p}$, $\bm{\tilde{Z}}^1_{a-p}$, $\bm{\tilde{Z}}^2_{a-p}$ are full square matrices of order $(2Q+1)(2R+1)$ which take the following form:
\begingroup\makeatletter\def\f@size{8}\check@mathfonts
\begin{align}
\bm{\tilde{Z}}^0_{a-p}=&
\begin{bmatrix}
\begin{smallmatrix}\bm{\tilde{M}}^0_{a-p,0}+\\-\bm{\tilde{K}}_{a-p,0}\bigg\{\bigg[(pk_{mx}+k_x)^2+(-Qk_{my}+k_y)^2\bigg]\cdot\\ \cdot\bigg[(ak_{mx}+k_x)^2+(-Qk_{my}+k_y)^2\bigg]+\\-(1-\nu)\bigg[([a-p]k_{mx})(-Qk_{my}+k_y)+\\-([-Q+Q]k_{my})(pk_{mx}+k_x)\bigg]^2\bigg\}\end{smallmatrix} & \dots & \begin{smallmatrix}      \bm{\tilde{M}}^0_{a-p,-2Q}+\\-\bm{\tilde{K}}_{a-p,-2Q}\bigg\{\bigg[(pk_{mx}+k_x)^2+(+Qk_{my}+k_y)^2\bigg]\cdot\\\cdot\bigg[(ak_{mx}+k_x)^2+(-Qk_{my}+k_y)^2\bigg]+\\-(1-\nu)\bigg[([a-p]k_{mx})(+Qk_{my}+k_y)+\\-([-Q-Q]k_{my})(pk_{mx}+k_x)\bigg]^2\bigg\}\end{smallmatrix}& \\
\vdots & \ddots & \vdots \\
\begin{smallmatrix}\bm{\tilde{M}}^0_{a-p,2Q}+\\-\bm{\tilde{K}}_{a-p,2Q}\bigg\{\bigg[(pk_{mx}+k_x)^2+(-Qk_{my}+k_y)^2\bigg]\cdot\\\cdot\bigg[(ak_{mx}+k_x)^2+(+Qk_{my}+k_y)^2\bigg]+\\-(1-\nu)\bigg[([a-p]k_{mx})(-Qk_{my}+k_y)+\\-([+Q+Q]k_{my})(pk_{mx}+k_x)\bigg]^2\bigg\}\end{smallmatrix} & \dots &  \begin{smallmatrix}\bm{\tilde{M}}^0_{a-p,0}+\\-\bm{\tilde{K}}_{a-p,0}\bigg\{\bigg[(pk_{mx}+k_x)^2+(+Qk_{my}+k_y)^2\bigg]\cdot\\\cdot\bigg[(ak_{mx}+k_x)^2+(+Qk_{my}+k_y)^2\bigg]+\\-(1-\nu)\bigg[([a-p]k_{mx})(+Qk_{my}+k_y)+\\-([+Q-Q]k_{my})(pk_{mx}+k_x)\bigg]^2\bigg\}\end{smallmatrix}
\end{bmatrix}\notag\\[4pt]
\bm{\tilde{Z}}^1_{a-p}=&
\begin{bmatrix}
\bm{\tilde{M}}^1_{a-p,0} & \dots & \bm{\tilde{M}}^1_{a-p,-2Q} \\
\vdots & \ddots & \vdots \\
\bm{\tilde{M}}^1_{a-p,2Q} & \dots & \bm{\tilde{M}}^1_{a-p,0}
\end{bmatrix} \notag\\[4pt]
\bm{\tilde{Z}}^2_{a-p}=&
\begin{bmatrix}
\bm{\tilde{M}}^2_{a-p,0} & \dots & \bm{\tilde{M}}^2_{a-p,-2Q} \\
\vdots & \ddots & \vdots \\
\bm{\tilde{M}}^2_{a-p,2Q} & \dots & \bm{\tilde{M}}^2_{a-p,0}
\end{bmatrix} \notag
\end{align}
\endgroup
which is further expanded for $p\in[-P,P]$ wave components:
\begin{equation}
\left[\bm{\tilde{L}}_0(k_x,k_y)+\bm{\tilde{L}}_1\omega+\bm{\tilde{L}}_2\omega^2\right]\bm{\tilde{w}}=0 \
\end{equation}
where $\bm{\tilde{L}}_0(k_x,k_y)$, $\bm{\tilde{L}}_1$, $\bm{\tilde{L}}_2$ are full square matrices of order\\ $(2P+1)(2Q+1)(2R+1)$:
\begingroup\makeatletter\def\f@size{8}\check@mathfonts
\begin{align}
\bm{\tilde{L}}_0(k_x,k_y)=&
\begin{bmatrix}
\bm{\tilde{Z}}^0_{0} & \dots & \bm{\tilde{Z}}^0_{-2P} \\
\vdots & \ddots & \vdots \\
\bm{\tilde{Z}}^0_{2P} & \dots & \bm{\tilde{Z}}^0_{0}
\end{bmatrix} \notag\\[0.2cm]
\bm{\tilde{L}}_1=&
\begin{bmatrix}
\bm{\tilde{Z}}^1_{0} & \dots & \bm{\tilde{Z}}^1_{-2P} \\
\vdots & \ddots & \vdots \\
\bm{\tilde{Z}}^1_{2P} & \dots & \bm{\tilde{Z}}^1_{0}
\end{bmatrix} \notag\\[0.2cm]
\bm{\tilde{L}}_2=&
\begin{bmatrix}
\bm{\tilde{Z}}^2_{0} & \dots & \bm{\tilde{Z}}^2_{-2P} \\
\vdots & \ddots & \vdots \\
\bm{\tilde{Z}}^2_{2P} & \dots & \bm{\tilde{Z}}^2_{0}
\end{bmatrix} \notag
\end{align}
\endgroup


\newpage
\bibliographystyle{elsarticle-num} 
\bibliography{Ref}
\end{document}